\documentclass[paper]{JHEP3}
\usepackage{cite}
\usepackage{amssymb,amsmath}
\usepackage{graphicx}
\graphicspath{{figures/}}

\def\Q2{\left(Q^{2}\right)}
\def\e{\epsilon}

\def\d{{\rm d}}
\def\JET{J}

\def\l({\left(}
\def\r){\right)}

\def\nf{N_{F}}
\def\kt{\tilde{k}}

\def\JET{J}
\def\e{\epsilon}

\def\d{\hbox{d}}

\def\sac{s_{ij}}
\def\sbc{s_{ik}}

\def\sabc{s_{ijk}}

\def\l{(}
\def\r{)}
\def\bl{\bigg(}
\def\br{\bigg)}

\title{Antenna subtraction at NNLO with hadronic initial states:
real-virtual initial-initial configurations}
\author{Thomas Gehrmann$^{a,b}$, Pier Francesco Monni$^a$\\
$^a$ Institut f\"ur Theoretische Physik, Universit\"at Z\"urich,
Wintherturerstrasse 190,\\CH-8057 Z\"urich, Switzerland\\
$^b$  Kavli Institute for Theoretical Physics, University of California,\\
Santa Barbara, CA 93106, USA}

\abstract{The antenna subtraction method handles 
real radiation contributions in higher order corrections to jet observables. 
The method is based on antenna functions, which encapsulate all 
unresolved radiation between a pair of hard radiator partons. To apply this method to 
compute hadron collider observables, 
initial-initial antenna functions with both radiators in the initial state are required. 
In view of extending the antenna subtraction method to next-to-next-to-leading order 
(NNLO) calculations at hadron colliders, we derive the one-loop initial-initial 
antenna functions in unintegrated and integrated form. }
\preprint{{ZU-TH 15/11, NSF-KITP-11-126}}

\begin{document}
%\maketitle
\allowdisplaybreaks

\section{Introduction}

Jet production observables are studied extensively at hadron colliders. Since 
the distribution of final state jets relates directly to the parton-level dynamics, jet 
observables can be used for precision studies of QCD~\cite{dissertori}, especially in 
view of determinations of the strong coupling constant and the parton 
distribution functions in the proton. 
Experimental measurements of these observables at the Tevatron~\cite{d0jet,cdfjet} 
attained an accuracy of a few per cent (or even better in certain kinematical ranges),
and first results from the LHC~\cite{atlasjet,cmsjet} 
already show the potential for precision jet physics.
Consequently, meaningful precision studies
must rely on theoretical predictions accurate to the same level. In
perturbative QCD, this precision usually requires corrections at
next-to-next-to-leading order (NNLO). 

NNLO calculations of observables with $n$ jets in the final state require
several ingredients: the two-loop corrected $n$-parton matrix elements, the
one-loop corrected $(n+1)$-parton matrix elements, and the tree-level
$(n+2)$-parton matrix elements. For most massless jet observables of
phenomenological interest, these matrix elements are available for some time 
already. 

The $(n+1)$-parton and $(n+2)$-parton matrix elements 
contribute to $n$ jet observables at NNLO if the extra partons are unresolved or
are clustered to form an
$n$-jet final state. Consequently, these extra partons are unconstrained in the 
soft and collinear regions, and  yield infrared divergences. 
In these cases, the infrared
singular parts of the matrix elements need to be extracted and integrated
over the phase space appropriate to the unresolved configuration
to make the infrared pole structure explicit. The single soft and collinear limits of 
one-loop matrix elements~\cite{onelstr,oneloopsoft,onelstr1,onelstr2,twolstr} and the  
double unresolved limits of tree-level matrix 
elements~\cite{audenigel,campbell,cg,campbellandother} are process-independent, 
and result in a factorization into an unresolved factor times a matrix element 
of lower multiplicity. 

To determine the contribution to NNLO jet observables from these configurations, one has to
find  subtraction terms which coincide with the full matrix element
and are still sufficiently simple to be integrated analytically in order
to cancel  their  infrared pole structure with the two-loop virtual contribution. 
Often starting from systematic
methods for subtraction at NLO~\cite{kunszt,cs,ant,nlosub},
several NNLO subtraction methods have been
proposed in the
literature~\cite{Kosower:2002su,nnlosub2,nnlosub3,nnlosub4,nnlosub5,nnlosub6,nnlosub7, nnlosub8}, and are
worked out
to a varying level of sophistication.

For observables with partons only in the final state,
an NNLO subtraction formalism, antenna subtraction,
 has been derived in~\cite{ourant}. The antenna subtraction formalism
constructs the subtraction terms from antenna functions. Each
antenna function encapsulates all singular limits due to the
emission of one or two unresolved partons between two colour-connected
hard radiator partons. This construction exploits the  universal factorization
of matrix elements and phase space in all unresolved limits.
The antenna functions are derived
systematically from physical matrix elements~\cite{our2j}. This formalism
has been applied in the derivation of NNLO corrections to three-jet
production in electron-positron annihilation~\cite{our3j,weinzierljet}
and related event shapes~\cite{ourevent,weinzierlevent}, which
were used subsequently in precision determinations of the strong coupling
constant~\cite{ouras,bechernew,hoang,davisonwebber,bethkeas,power}.
The formalism
can be extended to include parton showers at higher orders~\cite{antshower},
thereby offering a process-independent matching of fixed-order calculations
and logarithmic resummations~\cite{bechernew, GLM, hoang,resumall},
 which is done on a case-by-case
basis for individual observables~\cite{ourresum} up to now. The formalism
can be extended to include massive fermions~\cite{ritzmann}.

For processes
with initial-state partons, antenna subtraction has been fully worked out
only to NLO so far~\cite{hadant}. In this case, one encounters two new types
of antenna functions, initial-final antenna functions with one radiator parton
in the initial state, and initial-initial antenna functions with both
radiator partons in the initial state. The framework for the construction of 
NNLO antenna subtraction 
terms involving one or two partons in the initial state has been set up in~\cite{joao}
in the context of a proof-of-principle implementation of the contribution of the $gg\to 4g$ 
tree-level subprocess to di-jet production at hadron colliders. The initial-final  and 
initial-initial antenna functions appearing in the NNLO subtraction terms are 
obtained from crossing the final-final antennae. Their integration has to be performed 
over the appropriate phase space. In the case of the initial-final antennae, this 
has been accomplished in~\cite{gionata}. For the initial-initial 
tree-level double real radiation antenna functions, partial results have been 
obtained in~\cite{mathias}. It is the aim of the present paper to derive the setup for 
NNLO antenna subtraction for single unresolved singularities at one-loop and to 
compute the integrated one-loop initial-initial antenna functions required in this context.

Other approaches to perform NNLO calculations  of exclusive observables with initial
state partons are  the use of sector decomposition and a subtraction method based on the
transverse momentum structure of the final state.  The sector decomposition
algorithm~\cite{secdec}  analytically decomposes both phase space and loop integrals into
their  Laurent expansion in dimensional regularization, and  performs a subsequent
numerical  computation of the coefficients of this expansion. Using this formalism,  NNLO
results were  obtained for Higgs  production~\cite{babishiggs}     and vector boson
production~\cite{babisdy} at hadron colliders. Both reactions  were equally computed
independently~\cite{grazzinihiggs} using an NNLO subtraction  formalism exploiting the specific transverse
momentum  structure of these observables~\cite{nnlosub6}, which was also
applied most recently 
to compute NNLO corrections to associated $WH$ production~\cite{grazziniwh}.

This paper is structured as follows: in Section~\ref{sec:ant}, we
construct the subtraction terms required at NNLO
for initial-initial configurations with one unresolved parton at one loop. They require
one-loop $2\to 2$ antenna functions with two partons in the initial state
and one parton and one
off-shell neutral current in the final state. The analytic integration of the
initial-initial one-loop antenna functions is described in Section~\ref{sec:int}.
Finally, we conclude with an outlook in
Section~\ref{sec:conc}.

\section{Initial-intial antenna subtraction at NNLO}
\label{sec:ant}
Antenna subtraction of initial-initial configurations at NLO is derived in detail in~\cite{hadant}.
Subtraction terms with two hard partons in the initial state are built along the same
lines as in the final-final and initial-final case.
The NLO antenna subtraction term for an $m$-jet production process, 
to be convoluted with the
appropriate parton distribution functions for the initial state
partons,  for a configuration with the two hard emitters in the initial state 
(partons $i$ and $k$ with momenta $p_1$ and 
$p_2$), reads:
\begin{eqnarray}\label{eq:subii}
d\hat\sigma^{S,(II)}&=&{\cal N}\sum_{m+1}\d\Phi_{m+1}(k_1,\dots,k_{j-1},k_j,k_{j+1},\dots,k_{m+1};p_1,p_2)
  \,\frac{1}{S_{m+1}}\nonumber \\
&& \sum_{j}X^{0}_{ik,j}(p_1,p_2,k_j)
  \left|{\cal M}_m(\kt_1,\dots,\kt_{j-1},\kt_{j+1},\dots,\kt_{m+1}; x_1p_1,x_2p_2)\right|^2\nonumber\\
&&\times  J^{(m)}_{m}(\kt_1,\dots,\kt_{j-1},\kt_{j+1},\dots,\kt_{m+1})\,.  
\end{eqnarray}
All the momenta in the arguments of the reduced
matrix elements and the jet functions are redefined, which is a consequence of requiring 
the correct collinear factorization properties in both initial-state collinear limits. 
Its should be noted that the jet function $J^{(m)}_{m}$ (constructing $m$ jets from $m$ partons) 
requires all redefined momenta to be resolved. 
The two hard radiators are simply rescaled by factors 
$x_1$ and $x_2$ respectively. The spectator momenta are boosted 
by a Lorentz transformation onto the new set of 
momenta $\{\kt_l,\,l\neq j\}$. The mapping must be based on a factorization of the 
$(m+1)$-particle phase space, 
must satisfy overall momentum conservation and keep the mapped momenta 
on the mass shell. In this case, this turns out to severely restrict the
possible mappings. 

The tree-level antenna function $X^{0}_{ik,j}$ depends only on the incoming momenta $p_1,p_2$ and on the outgoing momentum $k_j$. It accounts for all 
singular configurations where parton $j$ is unresolved and colour-connected to 
partons $i$ (incoming with momentum $p_1$)  and $k$ (incoming with momentum $p_2$). 
The jet function $J^{(m)}_m$ and the reduced matrix element in (\ref{eq:subii}) depend
only on the redefined momenta. With a suitable factorization of the 
phase space~\cite{hadant}, one can perform the integration of the antenna function analytically. 

The factorization of the phase space is obtained by  requiring that the 
two mapped initial state momenta should be of the form
\begin{equation}
P_1=x_1p_1\,\qquad P_2=x_2p_2\, ,
\end{equation} 
so that 
\[\tilde q\equiv P_1+P_2\]
is in the beam axis. Since the vector component of 
$q\equiv p_1+p_2-k_j$ is in general not along the $p_1-p_2$ axis 
we need to boost all the momenta so that  $\tilde q = \Lambda q$ and $\tilde{k}_l = \Lambda k_l$
in order to restore momentum 
conservation. By requiring this boost to be only transverse, the phase space mapping 
is determined uniquely, resulting in the factorization
\begin{eqnarray}
\d\Phi_{m+1}(k_1,\dots,k_{m+1};p_1,p_2)&=&
\d\Phi_{m}(\kt_1,\dots,\kt_{j-1},\kt_{j+1},\dots,\kt_{m+1};x_1p_1,x_2p_2)
\nonumber\\
&&\times\delta(x_1-\hat{x}_1)\,\delta(x_2-\hat{x}_2)\,[\d k_j]\,\d x_1\,\d x_2\,.
\end{eqnarray}
where
$[\d k] = \d^d k \delta^{(+)}(k^2) /(2\pi)^{d-1}$ and 
\begin{eqnarray}\label{eq:mapii}
\hat{x}_1&=&\left(\frac{s_{12}-s_{j2}}{s_{12}}\,\frac{s_{12}-s_{1j}-s_{j2}}{s_{12}-s_{1j}}\right)^{\frac{1}{2}}\,,\nonumber \\
\hat{x}_2&=&\left(\frac{s_{12}-s_{1j}}{s_{12}}\,\frac{s_{12}-s_{1j}-s_{j2}}{s_{12}-s_{j2}}\right)^{\frac{1}{2}}\,.
\end{eqnarray}

Inserting the factorized expression for the phase space measure in
(\ref{eq:subii}), the subtraction terms can be integrated over
the antenna phase space. 
In the case of initial-initial subtraction terms, the antenna 
phase space is trivial: the two remaining Dirac delta functions
can be combined with the one particle phase space, such that there
are no integrals left. We define the initial-initial
integrated antenna functions as follows:
\begin{equation}
{\cal X}_{ik,j}(x_1,x_2)=\frac{1}{C(\epsilon)}\int
[\d k_j]\,x_1\,x_2\,\delta(x_1-\hat{x}_1)\,\delta(x_2-\hat{x}_2)\,X_{ik,j},
\end{equation}
where we introduced $C(\epsilon)=(4\pi)^\epsilon/(8\pi^2)e^{-\gamma_{E}\epsilon}$.\\
Substituting the one-particle phase space, and carrying out the
integrations over the Dirac delta functions, we have, 
\begin{eqnarray}
{\cal X}_{ik,j}(x_1,x_2)&=&({Q^2})^{-\epsilon}\frac{e^{\epsilon\gamma_E}}{\Gamma(1-\epsilon)}\,
{\cal J}(x_1,x_2)\,Q^2\,X_{ik,j}\,,
\label{eq:Xijk}
\end{eqnarray}
with $Q^2=q^2=(p_1+p_2-k_j)^2$. The Jacobian factor, ${\cal J}(x_1,x_2)$ is given by
\begin{equation}
\label{eq:jacobian}
{\cal
  J}(x_1,x_2)=\frac{x_1\,x_2\,(1+x_1\,x_2)}{(x_1+x_2)^2}\,(1-x_1)^{-\epsilon}(1-x_2)^{-\epsilon}\,
\left(\frac{(1+x_1)(1+x_2)}{(x_1+x_2)^2}\right)^{-\epsilon}\,,
\end{equation}
and the two-particle invariants are given by:
\begin{eqnarray}
s_{1j}=-s_{12}\frac{x_1\,(1-x_2^2)}{x_1+x_2}\,,\qquad
s_{j2}=-s_{12}\frac{x_2\,(1-x_1^2)}{x_1+x_2}\,.
\end{eqnarray} 
The integrated subtraction term is then,
\begin{eqnarray}
\d\hat{\sigma}^{S,(II)}&=&\sum_{m+1}\sum_{j}
\frac{{\cal N}}{S_{m+1}}\int\frac{\d x_1}{x_1}\frac{\d x_2}{x_2}\,
C(\epsilon)\,{\cal X}_{ik,j}(x_1,x_2)
\nonumber\\&&
\times \d\Phi_{m}(k_1,\dots,k_{j-1},k_{j+1},\dots,k_{m+1};x_1p_1,x_2p_2)
\nonumber\\&&
\times\left|{\cal
    M}_m(k_1,\dots,k_{j_1},k_{j+1},\dots,k_{m+1};x_1p_1,x_2p_2)\right|^2
\nonumber\\&&
\times J^{(m)}_{m}(k_1,\dots,k_{j-1},k_{j+1},\dots,k_{m+1})\,,
\end{eqnarray}
where we have relabeled all $\tilde{k}_i\to k_i$. 
The final step is to convolute this subtraction term with the parton
distribution functions of the initial state particles. The integrated version
of the subtraction pieces is then combined with the virtual and mass
factorization terms to yield a finite contribution when 
$\epsilon\rightarrow 0$. Recasting the convolutions appropriately, the
integrated subtraction term is
\begin{eqnarray}
\d\hat{{\sigma}}^{S,(II)}&=&\sum_{m+1}\sum_{j}
\frac{S_{m}}{S_{m+1}}\int\frac{\d\xi_1}{\xi_1}\int\frac{\d\xi_2}{\xi_2}
\int_{\xi_1}^{1}\frac{\d x_1}{x_1}
\,\int_{\xi_2}^{1}\frac{\d x_2}{x_2}
f_{i/1}\left(\frac{\xi_1}{x_2}\right)\,f_{k/2}\left(\frac{\xi_2}{x_2}\right)\nonumber\\
&&\times C(\epsilon)\,{\cal X}_{ik,j}(x_1,x_2)\,\d\hat{\sigma}^B(\xi_1 H_1,\xi_2 H_2)\,.
\end{eqnarray}
This convolution already has the appropriate
structure and combination with the virtual corrections and 
 mass factorization can be carried out explicitly leaving a finite
contribution. The remaining phase space integration, implicit in
the Born cross section, $\d\hat{\sigma}^B$, and the convolutions can
be safely evaluated numerically. 

At NNLO, two types of contributions to $m$-jet observables require
subtraction: the tree-level  ($m+2$)-parton matrix elements (where one or
two partons can become unresolved), and
the one-loop ($m+1$)-parton matrix elements (where one parton can become
unresolved). The corresponding
subtraction terms are denoted by  ${\rm d}\hat\sigma^{S}_{NNLO}$
and  $\d \hat\sigma^{VS,1}_{NNLO}$.
Antenna subtraction terms for the final-final~\cite{ourant} and initial-final~\cite{gionata}
cases have been derived previously.
In the initial-initial case, ${\rm d}\hat\sigma^{S}_{NNLO}$  was derived in~\cite{joao,mathias}. 
It contains subtraction terms for single unresolved limits (each containing a single 
three-particle antenna function $X_3^0$) and for double unresolved limits 
(containing four particle antenna functions $X_4^0$, products of three-particle antenna functions
($X_3^0\cdot X_3^0$) and soft large-angle correction terms ($S\cdot X_3^0$)). 

The integrand in the  ($m+1$)-parton channel consists, besides the  one-loop 
($m+1$)-parton matrix elements, of several contributions (independent of whether the 
radiators are in the initial or final state):
\begin{itemize}
\item[(a)] The integrated one-particle unresolved subtraction terms from the 
($m+2$)-parton channel, which cancel the 
explicit infrared poles of the virtual one-loop
$(m+1)$-parton matrix element.
\item[(b)] The virtual-unresolved subtraction term 
$\d \hat\sigma^{VS,1,b}_{NNLO}$ which subtracts all single unresolved limits from 
the virtual one-loop $(m+1)$-parton matrix element.
\item[(c)] Terms common to both above contributions,
which are oversubtracted. Each of these terms is formed by a product of an integrated and 
an unintegrated three-parton tree-level antenna function (${\cal X}_3^0 \cdot X_3^0$). These 
terms contain the full set of singly integrated ($X_3^0\cdot X_3^0$)-terms 
from ${\rm d}\hat\sigma^{S}_{NNLO}$, plus additional terms which 
must be further  integrated down to 
the $m$-parton channel. 
\item[(d)] The integrated soft large-angle correction terms (${\cal S}\cdot X_3^0$).  
\item[(e)] Terms arising from the mass factorization of the parton distribution functions at NLO. 
\end{itemize}
Unintegrated subtraction terms newly 
introduced in the $(m+1)$-parton channel have to be compensated by their integrated forms 
in the $m$-parton channel. The integration of contributions
 of type (b) in initial-initial kinematics
 is the main topic of this paper,
 they are already known for final-final~\cite{ourant} and initial-final~\cite{gionata}
 kinematics. 
 It should be noted that  integration of terms of type (c) does not require 
 any new integrals beyond the ${\cal X}_3^0$ already needed at NLO~\cite{hadant}. In
 particular, those terms obtained by integrating  ($X_3^0\cdot X_3^0$)
from ${\rm d}\hat\sigma^{S}_{NNLO}$ depend on the full set momenta of the $(m+1)$ partons
in a non-factorizable way, but are not integrated any further. Any additional terms of type (c) 
are chosen such that the ${\cal X}_3^0$ depends only on $m$-parton momenta obtained 
from the phase space mapping, such that the integration of  (${\cal X}_3^0 \cdot X_3^0$)
factorizes, involving only the known integral of $X_3^0$. Terms of type (d) will be dealt 
with elsewhere~\cite{joaonew}.

With radiator partons $i$ and $k$  in the initial state,
the contribution of type (b) reads:
\begin{eqnarray}
{\rm d}\hat\sigma_{NNLO}^{VS,1,b}
&= & {\cal N}\,\sum_{m+1}{\rm d}\Phi_{m+1}(k_{1},\ldots,k_{m+1};p_1,p_2)
\frac{1}{S_{{m+1}}} \nonumber \\
&\times& \,\sum_{j} \Bigg [X^0_{ik,j}\,
|{\cal M}^1_{m}(\tilde{k}_{1},\ldots,\tilde{k}_{j-1},\tilde{k}_{j+1},\ldots,
\tilde{k}_{m+1};x_1p_1,x_2p_2)|^2\,
\nonumber \\ && \hspace{3cm}\times
\JET_{m}^{(m)}(\tilde{k}_{1},\ldots,\tilde{k}_{j-1},\tilde{k}_{j+1},\ldots,\tilde{k}_{m+1})\;
\nonumber \\
&&\phantom{\sum_{j} }+\;X^1_{ik,j}\,
|{\cal M}_{m}(\tilde{k}_{1},\ldots,\tilde{k}_{j-1},\tilde{k}_{j+1},\ldots,
\tilde{k}_{m+1};x_1p_1,x_2p_2)|^2\,
\nonumber \\ && \hspace{3cm}\times
\JET_{m}^{(m)}(\tilde{k}_{1},\ldots,\tilde{k}_{j-1},\tilde{k}_{j+1},\ldots,\tilde{k}_{m+1})\;\Bigg
]\;,
\label{eq:subv2b}
\end{eqnarray}
In here, $X^1_{ik,j}$ denotes a one-loop three-parton initial-initial
antenna function, which is the only new ingredient. These antenna functions
can be obtained by crossing from their final-final counterparts, listed
in~\cite{ourant}, and have to be integrated over the appropriate
phase space according to (\ref{eq:Xijk}). 

\section{Integration of one-loop antenna functions}
\label{sec:int}
The one-loop antenna functions are derived from one-loop squared matrix
elements for all $2\to 2$  processes~\cite{our2j} 
obtained from 
$\gamma^*\to q\bar q g$ (quark-antiquark antenna functions), 
$\tilde{\chi} \to \tilde{g}gg$ and $\tilde{\chi}\to \tilde{g}q\bar q$ (quark-gluon
antenna functions) and $H\to ggg$, $H\to g q\bar q$ (gluon-gluon antenna 
functions) by crossing the off-shell current into the final state and two partons 
into the initial state. We denote a three-particle 
initial-initial antenna function with partons 
$(i,j)$  in the initial state and parton $k$ in the final state as $X^0_{ij,k}$ at tree-level 
and as $X^1_{ij,k}$ at one-loop. The tree-level and one-loop initial-initial antenna 
functions are summarized in Tables~\ref{tab:qqbar}--\ref{tab:gg}. The usual notation is used ,{\it i.e.} $X_{ij,k}^1$ for the leading colour ($N$) term, $\tilde{X}_{ij,k}^1$
for the subleading ($1/N$) term and $\hat{X}_{ij,k}^1$ for the $N_{F}$ part.
\TABLE[t]{
\renewcommand{\arraystretch}{1.2}
\begin{tabular}{p{5cm}p{3cm}p{4cm}}
\hline
Quark-antiquark initiated&tree level & one-loop \\
\hline
\underline{quark-quark}&&\\
$q\bar q \rightarrow g $&$A^{0}_{q\bar q,g}$&$A^{1}_{q\bar q,g}$, $\tilde{A}^{1}_{q\bar q,g}$,
$\hat{A}^{1}_{q\bar q,g}$\\[3mm]
\hline
\underline{quark-gluon}&&\\
$q \bar{q}^{\prime} \rightarrow {\bar q}^{\prime}$&
              $E^{0}_{q \bar{q}^{\prime}, \bar{q}^{\prime}}$&
              $E^{1}_{q \bar{q}^{\prime}, \bar{q}^{\prime}}$,
              $\tilde{E}^{1}_{q \bar{q}^{\prime}, \bar{q}^{\prime}}$,
              $\hat{E}^{1}_{q \bar{q}^{\prime}, \bar{q}^{\prime}}$\\
$q^\prime \bar{q}^{\prime} \rightarrow q $&
              $E^{0}_{q^\prime \bar{q}^{\prime}, q}$&
              $E^{1}_{q^\prime \bar{q}^{\prime}, q}$,
              $\tilde{E}^{1}_{q^\prime \bar{q}^{\prime}, q}$,
              $\hat{E}^{1}_{q^\prime \bar{q}^{\prime}, q}$\\[3mm]
\hline
\underline{gluon-gluon}&&\\
$q\bar{q}\rightarrow g $&$G^{0}_{q\bar{q},g}$&$G^{1}_{q\bar{q},g}$, $\tilde{G}^{1}_{q\bar{q},g}$, $\hat{G}^{1}_{q\bar{q},g}$\\[3mm]
\hline
\end{tabular}
\caption{List of tree level and one loop three-parton antenna functions for
the configurations with a quark-antiquark
system in the initial state.\label{tab:qqbar}}
}
\TABLE[t]{
\renewcommand{\arraystretch}{1.2}
\begin{tabular}{p{5cm}p{3cm}p{4cm}}
\hline
Quark-gluon initiated&tree level & one-loop \\
\hline
\underline{quark-quark}&&\\
$qg \rightarrow q $&$A^{0}_{qg,q}$&$A^{1}_{qg,q}$, $\tilde{A}^{1}_{qg,q}$,
$\hat{A}^{1}_{qg,q}$\\[3mm]
\hline
\underline{quark-gluon}&&\\
$q g \rightarrow g$&
              $D^{0}_{qg,g}$&
              $D^{1}_{qg,g}$,
              $\hat{D}^{1}_{qg,g}$\\[3mm]
\hline
\underline{gluon-gluon}&&\\
$q g \rightarrow q$&
              $G^{0}_{qg,q}$&
              $G^{1}_{qg,q}$,
              $\tilde{G}^{1}_{qg,q}$
              $\hat{G}^{1}_{qg,q}$\\[3mm]
\hline
\end{tabular}
\caption{List of tree level and one loop three-parton antenna functions for
the configurations with a quark-gluon
system in the initial state.\label{tab:qg}}
}
\TABLE[t]{
\renewcommand{\arraystretch}{1.2}
\begin{tabular}{p{5cm}p{3cm}p{4cm}}
\hline
Gluon-gluon initiated&tree level & one-loop \\
\hline
\underline{quark-gluon}&&\\
$g g \rightarrow q$&
              $D^{0}_{gg,q}$&
              $D^{1}_{gg,q}$,
              $\hat{D}^{1}_{gg,q}$\\[3mm]
\hline
\underline{gluon-gluon}&&\\
$g g \rightarrow g$&
              $F^{0}_{gg,g}$&
              $F^{1}_{gg,g}$,
              $\hat{F}^{1}_{gg,g}$\\[3mm]
\hline
\end{tabular}
\caption{List of tree level and one loop three-parton antenna functions for
the configurations with a gluon-gluon
system in the initial state.\label{tab:gg}}
}

We start from the unrenormalized one-loop squared three-parton matrix
elements (normalized to the 
corresponding two-parton matrix element and 
divided by a normalization factor $C(\e)$) relevant to
a particular antenna function, which we denote as $X^{1,U}_{ij,k} $. 
 The antenna
function is obtained after renormalization and subtraction of the corresponding
tree-level antenna function multiplied by the one-loop correction to the hard
radiator pair.
Renormalization of the one-loop antenna functions is always carried out in the $\overline{{\rm MS}}$-scheme
at fixed renormalization scale $\mu^2=Q^2$. It amounts to a renormalization of the strong
coupling constant and (in the case of the quark-gluon and gluon-gluon antenna functions)
to a renormalization of the effective operators used to couple an external current to
the partonic radiators. The relation between renormalized
and unrenormalized one-loop squared matrix elements is as follows:
\begin{eqnarray}
X^{1,R}_{ij,k} &=& X^{1,U}_{ij,k} -  \frac{b_0}{\e}\, X^0_{ij,k} - \frac{\eta_0}{\e}\, X^0_{ij,k}
\label{eq:renorm1}\,,\\
\tilde{X}^{1,R}_{ij,k} &=& \tilde{X}^{1,U}_{ij,k} \label{eq:renorm2}\,, \\
\hat{X}^{1,R}_{ij,k} &=& \hat{X}^{1,U}_{ij,k} -  \frac{b_{0,F}}{\e}\, X^0_{ij,k} - \frac{\eta_{0,F}}{\e}\, X^0_{ij,k}\,,
\end{eqnarray}
where
\begin{equation}
b_0 = \frac{11}{6}\;, \quad b_{0,F}= -\frac{1}{3}
\end{equation}
are the colour-ordered coefficients of the one-loop QCD $\beta$-function:
\begin{equation}
\beta_0 = b_0 N + b_{0,F} \nf\,.
\end{equation}
The renormalization constants for the effective operators are
\begin{eqnarray*}
 \eta_0 = 0\,, \quad \eta_{0,F}=0 \quad && \mbox{for $X=A$}\,,\\
 \eta_0 = b_0+\frac{3}{2} \,, \quad \eta_{0,F}=b_{0,F} \quad && \mbox{for $X=D,E$}\,,\\
 \eta_0 = 2\,b_0\,, \quad \eta_{0,F}=2\, b_{0,F} \quad && \mbox{for $X=F,G$}\,.
 \end{eqnarray*}

The one-loop antenna functions are obtained from the renormalized
one-loop  squared matrix elements by subtracting from them the product of
the tree-level antenna function with
the virtual one-loop hard radiator vertex
correction~\cite{onelstr,ourant}:
\begin{eqnarray}
X^1_{ij,k} &=& X^{1,R}_{ij,k}
- {\cal X}_2^1 \, X^0_{ij,k} \,,\\
\tilde{X}^1_{ij,k} &=& \tilde{X}^{1,R}_{ij,k}  - \tilde{{\cal X}}_2^1 \,
X^0_{ij,k}\,, \\
\hat{X}^1_{ij,k} &=& \hat{X}^{1,R}_{ij,k}
-  \hat{{\cal X}}_2^1\, X^0_{i,jk} \,.
\end{eqnarray}
 The one-loop corrections to the hard radiator vertex
are listed in~\cite{ourant}
 for ${\cal A}_2^1$, ${\cal D}_2^1$, $\hat{\cal D}_2^1$, ${\cal F}_2^1$
 and  $\hat{\cal F}_2^1$.
 From these, the remaining functions follow:
 \begin{eqnarray}
 \tilde{\cal A}_2^1&=& {\cal A}_2^1\,, \qquad \hat{\cal A}_2^1=0\,,\\
 \tilde{\cal D}_2^1&=& 0\,,\\
 {\cal E}_2^1 &=& {\cal D}_2^1\,\qquad
   \tilde{\cal E}_2^1= 0\,, \qquad \hat{\cal E}_2^1=\hat{\cal D}_2^1\,,\\
 \tilde{\cal F}_2^1&=& 0\,,\\
 {\cal G}_2^1 &=& {\cal F}_2^1\,\qquad
   \tilde{\cal G}_2^1= 0\,, \qquad \hat{\cal G}_2^1=\hat{\cal F}_2^1\,.
  \end{eqnarray}

The integrated 
forms of the single real radiation antenna functions are still differential 
in $x_1$ and $x_2$, such that no explicit integration has to be carried out.
However, endpoint singularities can occur in  either or both of these variables,
which have to be regularized dimensionally. This regularization is 
obtained from the $d$-dimensional integrated antenna functions
(\ref{eq:Xijk}) by expanding the product of the Jacobian factor and 
the antenna function in distributions using
\begin{equation}
(1-z)^{-1-\e}\,=\, -\frac{1}{\e} \, \delta(1-z) + \sum_{n}
\frac{(-\e)^{n}}{n!}\mathcal{D}_{n}(1-z)\,,
\label{eq:dist}
\end{equation}
with
$$\mathcal{D}_{n}(1-z)=\left(\frac{\ln^{n}\left(1-z\right)}{1-z}\right)_{+}\,.$$

The expansion is straightforward for the tree-level antenna functions~\cite{hadant}, which 
contain only rational factors in the invariants
\begin{equation}
s_{12} = \frac{q^2}{x_1x_2}\,, \qquad 
s_{1j}=-q^2\frac{1-x_2^2}{x_2(x_1+x_2)}\,,\qquad
s_{2j}=-q^2\frac{1-x_1^2}{x_1(x_1+x_2)}\,,
\end{equation}
where we 
denote the pair of initial state partons with  indices 
$(1,2)$ and we refer to the unresolved one with the index $j$. 
The one-loop antenna functions contain logarithms and polylogarithms, 
yielding branch-cuts at the kinematical endpoints, which forbid 
a direct expansion in distributions. Instead, we start from the 
unintegrated expression in terms of one-loop master integrals. Only two 
types of master integrals appear: the one-loop bubble
\begin{equation}
{\rm Bub}(p^2)= \left[\frac{(4\pi)^{\e}}{16\pi^2}\frac{ \Gamma (1+\e)
    \Gamma^2 (1-\e)}{ \Gamma (1-2\e)} \right]  \; \frac{i}{\e\,(1-2\e)}
\left( -p^2\right)^{-\e} \equiv A_{2,LO} 
\left( -p^2\right)^{-\e} \;.
\end{equation}
and the general one-loop box with one off-shell leg
\begin{eqnarray}
\lefteqn{{\rm Box}(\sac,\sbc)   =   \frac{2(1-2\e)}{\e} A_{2,LO}
\frac{1}{\sac\sbc}} \nonumber \\ 
 & & \hspace{0.46cm}
\Bigg[ \left(\frac{\sac \sbc}{\sac - \sabc}\right)^{-\e}
 \,_2F_1\left(-\e, -\e; 1-\e; \frac{\sabc - \sac - \sbc}{\sabc -\sac}
 \right)\nonumber \\ 
 & & \hspace{0.4cm}
+ \left(\frac{\sac \sbc}{\sbc - \sabc}\right)^{-\e}
 \,_2F_1\left(-\e,-\e;1-\e; \frac{\sabc - \sac - \sbc}{\sabc -\sbc}
 \right) \nonumber \\
& & \hspace{0.4cm} - \left( \frac{-\sabc \sac \sbc}{(\sac - \sabc)
   (\sbc - \sabc)} 
  \right)^{-\e} \,_2F_1\left(-\e, -\e; 1-\e; 
\frac{\sabc (\sabc - \sac - \sbc)}{(\sabc -\sac)(\sabc - \sbc)}\right) \Bigg]
\;.
\nonumber \\
\end{eqnarray}
Both integrals appear in all kinematical crossings. The expansion of the 
terms involving the bubble integral is trivial, while
the one-loop box
${\rm Box}(s_{ij},s_{ik})$ contains the rational factor $1/(s_{ij}\, s_{ik})$ and 
appears in the unintegrated antenna functions with further 
rational prefactors. In terms of the expansion in distributions, one 
has to distinguish three prefactors: $1$,  $s_{ij}/s_{jk}$ and $s_{ik}/s_{jk}$. 

In order to analyse the initial-initial kinematical configuration we limit ourselves to the case
\begin{align}
s_{12}>q^2>0,\qquad s_{1j}<0,\qquad s_{2j}<0,
\label{eq:kinreg}
\end{align}
and we study all the possible crossings of the master integrals. For each of the three crossings of the box integral,  
expansions have to be derived for each of the three prefactors mentioned above. 

These expansions proceed by analytic continuation of the hypergeometric 
functions in the box master integrals to the appropriate region of analyticity,
and requiring that the limit $x_i\to 1$ does not result in an argument of the 
hypergeometric function equal to 1 or infinity (avoiding the branch-cut). In this situation, 
(\ref{eq:dist}) can be applied safely to the coefficients of the 
hypergeometric function. 

Using the following notation
\begin{align}
\label{eq:boxnotation}
{\rm M_{box}}\left(s_{ij},s_{ik},\frac{s_{lm}}{s_{pq}}\right) =& \frac{2}{C(\epsilon)} {\cal J}(x_1,x_2) \frac{s_{lm}}{s_{pq}} {\cal R}\left({\rm Box}(s_{ij},s_{ik})\right),\\
\label{eq:bubblenotation}
{\rm M_{bub}}(p^2) =& \frac{2}{C(\epsilon)} {\cal J}(x_1,x_2) \mathcal{R}\left({\rm Bub}(p^2)\right),
\end{align}
where $\mathcal{R}$ selects the real part, we list the master integrals
 with the relevant prefactors in Appendix~\ref{app:A}.

The resulting expressions for the integrated antenna functions 
${\cal X}^1_{ij,k}$ are very lengthy, such that we only quote one of them, $\hat{{\cal D}}_{qg,g}$,
in Appendix~\ref{app:B} as an example. 
Analytic expressions 
for all of them, 
as well as for the tree-level  antenna functions ${\cal X}^0_{ij,k}$ 
expanded through to ${\cal O}(\e^2)$ are attached with the arXiv-submission of this article. 

\section{Conclusions}
\label{sec:conc}
In this paper, we extended the antenna subtraction formalism to handle 
single unresolved radiation at one loop for processes with two partons in the 
initial state, as required for hadron collider cross sections at NNLO accuracy. The corresponding 
virtual unresolved subtraction terms consist of tree-level and one-loop antenna functions
with both radiators in the initial state. These initial-initial antenna functions  are required 
in unintegrated and integrated form. The unintegrated 
initial-initial antenna functions are obtained 
straightforwardly
from analytic continuation of the corresponding final-final antennae. The integration
of  the one-loop 
antennae over the phase space relevant to the initial-initial configurations requires 
an expansion in distributions around the kinematical endpoints of the two initial-state 
momentum fractions, which we performed for all relevant master integrals. 

Using the results of this paper
in combination with the one-loop antenna functions in the 
initial-final~\cite{gionata} and final-final~\cite{ourant} case, 
the NNLO subtraction terms for the one-loop 
$(n+1)$-parton contribution to $n$-jet observables
at hadron colliders can be constructed and implemented. 
To accomplish a full NNLO description of $n$-jet observables
at hadron colliders, subtraction terms for
the double real radiation $(n+2)$-parton contribution are equally needed. The 
construction of these subtraction terms in the case of hadronic collisions 
has been described in~\cite{joao}. A large fraction of the integrated antenna functions 
have been derived already~\cite{mathias}. Once the full set of integrated 
double real radiation antenna functions
is completed for the initial-initial case, the antenna subtraction method 
can be applied to the computation of NNLO corrections to jet production at 
hadron colliders.

\section*{Acknowledgements}
TG would like to thank the Kavli Institute for 
Theoretical Physics (KITP) at UC Santa Barbara for hospitality while this 
work was completed. 
This research is supported in part by
the Swiss National Science Foundation (SNF) under contract
200020-126691,  by the European Commission through the 
``LHCPhenoNet" Initial Training Network PITN-GA-2010-264564 and 
by the National Science Foundation under grant NSF PHY05-51164.

\appendix
\section{Master integrals in initital-initial configuration}
\label{app:A}
The initial-initial one-loop antenna functions can be expressed by the bubble and 
box master integrals defined in (\ref{eq:boxnotation}--\ref{eq:bubblenotation}). They 
require analytic continuation to the appropriate kinematical region~(\ref{eq:kinreg}).

The analytic continuation of the general bubble integral is straightforward
\begin{align}
{\rm M_{bub}}(p^2) = \frac{2}{C(\epsilon)} {\cal J}(x_1,x_2)A_{2,LO}
{\cal R}\left( -p^2-i\delta\right)^{-\e},
\end{align}
for $p^2>0$. The 
box integrals read:
\small
\begin{align}
&(Q^2)^{2+\epsilon}{\rm M_{box}}(s_{1j},s_{2j},1) = 
%#i####
-\frac{1}{8\epsilon^4} \delta(1-x_{1}) \delta(1-x_{2})-\frac{1}{2\epsilon^3} \delta(1-x_{2}) (1+x_{1}-\mathcal{D}_{0}(1-x_{1}))+\notag\\
&\frac{5 \pi ^2}{96\epsilon^2}\delta(1-x_{1}) \delta(1-x_{2})+\frac{1}{2\epsilon^2} \bigg((2+2 x_{1}-\mathcal{D}_{0}(1-x_{1})) \mathcal{D}_{0}(1-x_{2})-\notag\\
&\delta(1-x_{2}) \bigg(2\mathcal{D}_{1}(1-x_{1})-2 (1+x_{1}) \log\l1-x_{1}\r-\frac{x_{1}^2}{1-x_{1}} \log\bl\frac{1}{4} x_{1} (1+x_{1})^2\br\bigg)\bigg)-\notag\\
&\frac{1}{2\epsilon^2}\frac{1}{(1+x_{1}) (1+x_{2})}\left((1+x_{2})^2+2 x_{1} (1+x_{2})^2+x_{1}^2 (1+2 x_{2} (1+x_{2}))\right)-\notag\\
&\frac{1}{24\epsilon}\bigg(24 \bigg(2 (1+x_{1}) \mathcal{D}_{1}(1-x_{2})+\mathcal{D}_{0}(1-x_{2}) \bigg(-2 \mathcal{D}_{1}(1-x_{1})+\notag\\
&2 (1+x_{1}) \log\l1-x_{1}\r+\frac{x_{1}^2}{1-x_{1}} \log\bl\frac{1}{4} x_{1} (1+x_{1})^2\br\bigg)\bigg)+\delta(1-x_{2}) \bigg(-5 \pi^2 (1+x_{1})+\notag\\
&5 \pi ^2\mathcal{D}_{0}(1-x_{1})-24\mathcal{D}_{2}(1-x_{1})+\frac{6}{1-x_{1}} \bigg(-4 x_{1}^2 \log\l4(1-x_{1})\r \log\l1-x_{1}\r+\notag\\
&4\log^{2}\l1-x_{1}\r+4 x_{1}^2 \log\l1-x_{1}\r \log\bl(1+x_{1})^2\br+x_{1}^2 \bigg(\log^{2}\bl\frac{4}{(1+x_{1})^2}\br-2\bigg(\log^{2}\bl\frac{1}{4}(1+x_{1})^2\br+\notag\\
&\log\l x_{1}\r \log\bl\frac{(1+x_{1})^2}{4(1-x_{1})}\br\bigg)\bigg)\bigg)-12 \frac{x_{1}^2}{1-x_{1}}{\rm Li}_{2}\l1-x_{1}\r-7\delta(1-x_{1})\zeta_{3}\bigg)\bigg)+\notag\\
&\frac{1}{\epsilon}\bigg(\frac{2}{(1+x_{1}) (1+x_{2})}((1+x_{2})^2+2 x_{1} (1+x_{2})^2+x_{1}^2 (1+2 x_{2} (1+x_{2})))\log\l1-x_{1}\r+\notag\\
&\frac{x_{1}^2}{(1-x_{1})(1-x_{2})} \log\bl4 x_{1} (1+x_{1})^2\br-\frac{2 x_{1}^2 x_{2}^2 (1+x_{1} x_{2})}{(1-x_{1}^2)(1-x_{2}^2)} \log\bl\frac{x_{1} (1+x_{1} x_{2})}{x_{1}+x_{2}}\br-\notag\\
&\frac{(x_{1}^2+x_{2}^2)}{(1-x_{1})(1-x_{2})}\log\l4\r-\frac{4 x_{1}^2 x_{2}^2 (1+x_{1} x_{2})}{(1-x_{1}^2)(1-x_{2}^2)} \log\bl\frac{(x_{1}+x_{2})^2}{(1+x_{1})^2}\br-\notag\\
&\frac{2 x_{1}^2 x_{2}^2 (1+x_{1} x_{2})}{(1-x_{1}^2)(1-x_{2}^2)} \log\bl\frac{(1+x_{1})^2}{(x_{1}+x_{2}) (1+x_{1} x_{2})}\br\bigg)-\notag\\
&\frac{1}{3840}\bigg(-13\pi ^4\delta(1-x_{1}) \delta(1-x_{2})-160 \bigg(-48\bigg(\mathcal{D}_{1}(1-x_{1}) \mathcal{D}_{1}(1-x_{2})-\notag\\
&(1+x_{2}) \mathcal{D}_{2}(1-x_{1})-\mathcal{D}_{1}(1-x_{2}) \bigg(2(1+x_{1}) \log\l1-x_{1}\r+\frac{x_{1}^2}{1-x_{1}} \log\bl\frac{1}{4} x_{1} (1+x_{1})^2\br\bigg)\bigg)+\notag\\
&\mathcal{D}_{0}(1-x_{2}) \bigg(5 \pi ^2 \mathcal{D}_{0}(1-x_{1})+2 \bigg(-5 \pi ^2 (1+x_{1})+24 \frac{\log^2\l1-x_{1}\r}{1-x_{1}}+\notag\\
&6\bigg(-4\mathcal{D}_{2}(1-x_{1})+\frac{x_{1}^2}{1-x_{1}}\bigg(\log^2\bl\frac{4}{(1+x_{1})^2}\br+4\log\l1-x_{1}\r \log\bl\frac{(1+x_{1})^2}{4 (1-x_{1})}\br-\notag\\
&2\bigg(\log^2\bl\frac{1}{4}(1+x_{1})^2\br+\log\l x_{1}\r \log\bl\frac{(1+x_{1})^2}{4(1-x_{1})}\br\bigg)-2{\rm Li}_{2}\l1-x_{1}\r\bigg)\bigg)\bigg)\bigg)+\notag\\
&\delta(1-x_{2}) \bigg(10 \pi ^2\mathcal{D}_{1}(1-x_{1})-16\mathcal{D}_{3}(1-x_{1})-10 \pi ^2 \log\l1-x_{1}\r(1+x_{1})+\notag\\
&16 \log^3\l1-x_{1}\r(1+x_{1})+\frac{x_{1}^2}{1-x_{1}}\bigg(-\pi ^2\log\l x_{1}\r+24\log^2\l1-x_{1}\r \log\bl\frac{1}{4 (1+x_{1})^2}\br-\notag\\
&5 \pi ^2\log\bl\frac{(1+x_{1})^2}{4}\br+12\log\l1-x_{1}\r \log^2\bl\frac{4}{(1+x_{1})^2}\br+2\log^3\bl\frac{4}{(1+x_{1})^2}\br-\notag\\
&24\log\l1-x_{1}\r \log^2\bl\frac{1}{4} (1+x_{1})^2\br+4\log^3\bl\frac{1}{4} (1+x_{1})^2\br+48\log^2\l1-x_{1}\r\log\bl(1+x_{1})^2\br+\notag\\
&6\log\l x_{1}\r \log^2\bl\frac{(1+x_{1})^2}{4(1-x_{1})}\br+12\log\bl\frac{(1+x_{1})^2}{4(1-x_{1})}\br{\rm Li}_{2}\l1-x_{1}\r-12{\rm Li}_{3}\l1-x_{1}\r\bigg)+\notag\\
&28(1+x_{1}-\mathcal{D}_{0}(1-x_{1}))\zeta_{3}\bigg)\bigg)\bigg)-\frac{x_{1}^2}{2(1-x_{1})(1-x_{2})} \log^2\bl\frac{4}{(1+x_{1})^2}\br-\notag\\
&\frac{2}{(1+x_{1}) (1+x_{2})} \left((1+x_{2})^2+2 x_{1} (1+x_{2})^2+x_{1}^2 (1+2 x_{2} (1+x_{2}))\right)\log^2\l1-x_{1}\r+\notag\\
&\frac{x_{1}^2}{(1-x_{1})(1-x_{2})}\left( \log\l x_{1}\r \log\bl\frac{(1+x_{1})^2}{4(1-x_{1})}\br+2\log\bl\frac{4}{(1+x_{1})^2}\br \left(-\log\l1-x_{1}\r-\log\l1-x_{2}\r\right)\right)+\notag\\
&\frac{8 x_{1}^2 x_{2}^2 (1+x_{1} x_{2}) \log\l1-x_{1}\r}{(1-x_{1}^2)(1-x_{2}^2)}\log\bl\frac{(x_{1}+x_{2})^4}{(1+x_{1})^2 (1+x_{2})^2}\br-\frac{x_{1}^2 x_{2}^2 (1+x_{1} x_{2})}{(1-x_{1}^2)(1-x_{2}^2)}\log^2\bl\frac{(x_{1}+x_{2})^4}{(1+x_{1})^2 (1+x_{2})^2}\br-\notag\\
&\frac{2 }{(1+x_{1}) (1+x_{2})}\left((1+x_{2})^2+2 x_{1} (1+x_{2})^2+x_{1}^2 (1+2 x_{2} (1+x_{2}))\right)\log\l1-x_{2}\r \log\l1-x_{1}\r +\notag\\
&\frac{\log\l1-x_{1}\r}{(1-x_{1})(1-x_{2})} \bigg(-2 x_{2}^2 \log\l x_{2}\r+4((x_{1}^2+x_{2}^2) \log\l4\r-2 x_{1}^2 \log\l1+x_{1}\r-2 x_{2}^2 \log\l1+x_{2}\r)+\notag\\
&4 \frac{x_{1}^2 x_{2}^2(1+x_{1}x_{2})}{(1+x_{1})(1+x_{2})} \log\bl\frac{(1+x_{1})^2 (1+x_{2})^2}{(x_{1}+x_{2})^2 (1+x_{1} x_{2})^2}\br\bigg)+\bigg(\frac{4 x_{1}^2 x_{2}^2 (1+x_{1} x_{2})}{(1-x_{1}^2)(1-x_{2}^2)} \log\l1-x_{2}\r-\notag\\
&\frac{2 x_{1}^2 x_{2}^2 (1+x_{1} x_{2}) }{(1-x_{1}^2)(1-x_{2}^2)}\log\bl\frac{x_{2} (x_{1}+x_{2})^3}{(1-x_{1}^2)(1+x_{2})^2}\br\bigg) \log\bl\frac{x_{1} (1+x_{1} x_{2})}{x_{1}+x_{2}}\br+\notag\\
&\frac{1}{24(1-x_{1}^2)(1-x_{2}^2)}\bigg(\pi ^2 \big(5 (1-x_{2})(1+x_{1})(1+x_{2})^2+x_{1}^3(-5-5 x_{2}+6 x_{2}^3)+\notag\\
&x_{1}^2 (-5+x_{2} (-5+6 x_{2}))\big)+12 (1+x_{1}) (1+x_{2})(x_{1}^2+x_{2}^2)\log^2\l4\r+\notag\\
&48 (1+x_{1}) (1+x_{2}) \bigg(x_{1}^2 \log\bl\frac{1+x_{1}}{4}\br\log\l1+x_{1}\r+x_{2}^2 \log\bl\frac{1+x_{2}}{4}\br \log\l1+x_{2}\r\bigg)+\notag\\
&12 x_{1}^2 x_{2}^2 (1+x_{1} x_{2}) \log^2\bl\frac{(1+x_{1})^2 (1+x_{2})^2}{(x_{1}+x_{2})^2 (1+x_{1} x_{2})^2}\br+12 x_{1}^2 (1+x_{1}) (1+x_{2}) {\rm Li}_{2}\l1-x_{1}\r+\notag\\
&12 x_{2}^2 \bigg((1+x_{1}) (1+x_{2}) {\rm Li}_{2}\l1-x_{2}\r+2 x_{1}^2 (1+x_{1} x_{2}) \bigg({\rm Li}_{2}\l\frac{(x_{1}+x_{2})^2}{(1+x_{1} x_{2})^2}\r-\notag\\
&{\rm Li}_{2}\l\frac{x_{2}-x_{1}^2 x_{2}}{x_{1}+x_{2}}\r-{\rm Li}_{2}\l\frac{x_{1}-x_{1} x_{2}^2}{x_{1}+x_{2}}\r\bigg)\bigg)\bigg)
%#o####
+\{x_{1}\leftrightarrow x_{2}\}+\mathcal{O}(\epsilon),
\end{align}

\;
\;

\small
\begin{align}
&(Q^2)^{2+\epsilon}{\rm M_{box}}(s_{12},s_{1j},1) = 
%#i####
-\frac{1}{2\epsilon^3} x_{1}^2 \delta(1-x_{2})+\frac{1}{2\epsilon^2} x_{1}^2 \bigg(2 \mathcal{D}_{0}(1-x_{2})+\delta(1-x_{2}) \log\bl\frac{4(1-x_{1})}{x_{1} (1+x_{1})^2}\br\bigg)+\notag\\
&\frac{x_{1}^2}{\epsilon^2}\bigg(-2+\frac{1}{1+x_{2}}-\frac{2 x_{1} x_{2}^2}{x_{1}+x_{2}}\bigg)+\frac{1}{24\epsilon} x_{1}^2 \bigg(24 \bigg(-2 \mathcal{D}_{1}(1-x_{2})+\mathcal{D}_{0}(1-x_{2}) \log\bl\frac{x_{1} (1+x_{1})^2}{4 (1-x_{1})}\br\bigg)\bigg)+\notag\\
&\frac{1}{\epsilon}\bigg(x_{1}^2 \big(x_{2}+2 x_{2}^2+x_{1}(1+2 x_{2}(1+x_{2}+x_{2}^2))\big) \frac{\log\l1-x_{1}\r}{(1+x_{2}) (x_{1}+x_{2})}-\frac{x_{1}^2 \log\l x_{1}\r}{1-x_{2}}+\notag\\
&2 x_{1}^2 \big(x_{2}+2 x_{2}^2+x_{1}(1+2 x_{2}(1+x_{2}+x_{2}^2))\big)\frac{\log\l1-x_{2}\r}{(1+x_{2}) (x_{1}+x_{2})}-\frac{x_{1}^2}{1-x_{2}} \bigg(\log\bl\frac{(1+x_{1})^2}{4}\br-\notag\\
&2 \frac{x_{2}^3 (1+x_{1} x_{2})}{(x_{1}+x_{2})(1+x_{2})} \log\bl\frac{x_{1} x_{2}^2 (x_{1}+x_{2})^3}{(1+x_{1}) (1+x_{2})^2}\br\bigg)\bigg)+\notag\\
&\delta(1-x_{2}) \frac{1}{24\epsilon} x_{1}^2 \bigg(\pi ^2-6 \log\bl\frac{4(1-x_{1})}{(1+x_{1})^2}\br \log\bl\frac{4(1-x_{1})}{x_{1}^2 (1+x_{1})^2}\br+12{\rm Li}_{2}\l1-x_{1}\r\bigg)+\notag\\
&\frac{1}{24} x_{1}^2 \bigg(48 \bigg(\mathcal{D}_{2}(1-x_{2})+\mathcal{D}_{1}(1-x_{2}) \log\bl\frac{4(1-x_{1})}{x_{1}(1+x_{1})^2}\br\bigg)-2 \mathcal{D}_{0}(1-x_{2})\bigg(\pi ^2+12\log\l x_{1}\r \log\bl\frac{4(1-x_{1})}{(1+x_{1})^2}\br-\notag\\
&6\log^2\bl\frac{4(1-x_{1})}{(1+x_{1})^2}\br+12 {\rm Li}_{2}\l1-x_{1}\r\bigg)+\delta(1-x_{2}) \bigg(2 \log\l1-x_{1}\r \bigg(\log\l1-x_{1}\r \log\bl\frac{64(1-x_{1})}{(1+x_{1})^6}\br-\notag\\
&3 \log\l x_{1}\r\log\bl\frac{16(1-x_{1})}{(1+x_{1})^4}\br\bigg)-\log\bl\frac{1-x_{1}}{x_{1}}\br \bigg(\pi ^2-6 \log^2\l4\r+6 \log\bl\frac{16}{(1+x_{1})^2}\br \log\bl(1+x_{1})^2\br\bigg)+\notag\\
&\log\bl\frac{4}{(1+x_{1})^2}\br \bigg(-\pi ^2+2 \log^2\l4\r+2 \log\bl\frac{1}{16} (1+x_{1})^2\br \log\bl(1+x_{1})^2\br\bigg)+\notag\\
&12\log\bl\frac{(1+x_{1})^2}{4(1-x_{1})}\br {\rm Li}_{2}\l1-x_{1}\r+12 {\rm Li}_{3}\l1-x_{1}\r+28 \zeta_{3}\bigg)\bigg)-\frac{2 x_{1}^2 }{1-x_{2}}\log\l x_{1}\r \log\bl\frac{1+x_{1}}{2}\br-\notag\\
&x_{1}^2 \big(x_{2}+2 x_{2}^2+x_{1}(1+2 x_{2}(1+x_{2}+x_{2}^2))\big) \frac{\log^2\l1-x_{1}\r}{2 (1+x_{2}) (x_{1}+x_{2})}-\notag\\
&2 x_{1}^2 \big(x_{2}+2 x_{2}^2+x_{1}(1+2 x_{2}(1+x_{2}+x_{2}^2))\big) \frac{\log^2\l1-x_{2}\r}{(1+x_{2}) (x_{1}+x_{2})}+\log\l1-x_{1}\r \bigg(\frac{x_{1}^2 \log\l x_{1}\r}{1-x_{2}}-\notag\\
&\frac{x_{1}^2}{(x_{1}+x_{2})(1-x_{2}^2)} \bigg((1+x_{2}) (x_{1}+x_{2}) \log\l4\r-2 (1+x_{2}) (x_{1}+x_{2}) \log\l1+x_{1}\r+\notag\\
&2 x_{2}^3 (1+x_{1} x_{2}) \log\bl\frac{x_{1} x_{2}^2 (x_{1}+x_{2})^3}{(1+x_{1}) (1+x_{2})^2}\br\bigg)\bigg)+\log\l1-x_{2}\r \bigg(\frac{2 x_{1}^2 \log\l x_{1}\r}{1-x_{2}}-2 x_{1}^2 \big(x_{2}+2 x_{2}^2+\notag\\
&x_{1}(1+2 x_{2}(1+x_{2}+x_{2}^2))\big) \frac{\log\l1-x_{1}\r}{(1+x_{2}) (x_{1}+x_{2})}-\frac{2 x_{1}^2}{(x_{1}+x_{2})(1-x_{2}^2)} \bigg((1+x_{2}) (x_{1}+x_{2}) \log\l4\r-\notag\\
&2 (1+x_{2}) (x_{1}+x_{2}) \log\l1+x_{1}\r+2 x_{2}^3 (1+x_{1} x_{2}) \log\bl\frac{x_{1} x_{2}^2 (x_{1}+x_{2})^3}{(1+x_{1}) (1+x_{2})^2}\br\bigg)\bigg)-\notag\\
&\frac{1}{12 (x_{1}+x_{2})(1-x_{2}^2)}x_{1}^2 \bigg(-\pi ^2(1-x_{2}) \big(x_{2}+2 x_{2}^2+x_{1}(1+2 x_{2}(1+x_{2}+x_{2}^2))\big)+\notag\\
&6 (1+x_{2}) (x_{1}+x_{2}) \log^2\l4\r+24 (1+x_{2}) (x_{1}+x_{2}) \log\bl\frac{1+x_{1}}{4}\br \log\l1+x_{1}\r-\notag\\
&12 x_{2}^3 (1+x_{1} x_{2}) \bigg(\log^2\bl\frac{x_{2} (x_{1}+x_{2})^3}{(1+x_{1}) (1+x_{2})^2}\br+\log^2\bl\frac{x_{2} (x_{1}+x_{2})^3 (1-x_{1} x_{2})}{(1+x_{1}) (1+x_{2})^2}\br+\notag\\
&2 \log\bl\frac{x_{2} (x_{1}+x_{2})^3}{(1+x_{1}) (1+x_{2})^2}\br \log\bl\frac{x_{1} (1+x_{1} x_{2})}{x_{1}+x_{2}}\br-\log^2\bl\frac{(x_{1}+x_{2})^2(1-x_{1}^2 x_{2}^2)}{(1+x_{1}) (1+x_{2})^2}\br\bigg)-\notag\\
&12 (1+x_{2}) (x_{1}+x_{2}) {\rm Li}_{2}\l1-x_{1}\r+24 x_{2}^3 (1+x_{1} x_{2}) \bigg(-{\rm Li}_{2}\bl\frac{(1-x_{1}^2)x_{2}}{(x_{1}+x_{2})(1-x_{1} x_{2})}\br+\notag\\
&{\rm Li}_{2}\bl\frac{x_{2}-x_{1}^2 x_{2}}{x_{1}+x_{2}}\br+{\rm Li}_{2}\bl\frac{(1-x_{1}^2)x_{2}^2}{1-x_{1}^2 x_{2}^2}\br\bigg)\bigg)
%#o####
+\mathcal{O}(\epsilon),
\end{align}

\;
\;

\small
\begin{align}
&(Q^2)^{2+\epsilon}{\rm M_{box}}(s_{12},s_{1j},\frac{s_{12}}{s_{2j}}) = 
%#i####
-\frac{1}{2\epsilon^4} \delta(1-x_{1})\delta(1-x_{2})-\frac{1}{2\epsilon^3} \delta(1-x_{2}) (1+x_{1}-\mathcal{D}_{0}(1-x_{1}))+\notag\\
&\frac{1}{\epsilon^3}\delta(1-x_{1}) (-1-x_{2}+\mathcal{D}_{0}(1-x_{2}))+\frac{1}{\epsilon^2}\mathcal{D}_{0}(1-x_{1}) (1+x_{2}-\mathcal{D}_{0}(1-x_{2}))+\frac{1}{\epsilon^2}(1+x_{1}) \mathcal{D}_{0}(1-x_{2})+\notag\\
&\frac{1}{2\epsilon^2}\delta(1-x_{2}) \bigg(-\mathcal{D}_{1}(1-x_{1})+\frac{\log\l1-x_{1}\r}{1-x_{1}}+\frac{x_{1}^2}{1-x_{1}} \log\bl\frac{x_{1} (1+x_{1})^2}{4 (1-x_{1})}\br\bigg)-\notag\\
&\frac{1}{24\epsilon^2}\delta(1-x_{1}) \bigg(-\pi ^2\delta(1-x_{2})-24 \bigg(-2\mathcal{D}_{1}(1-x_{2})+2(1+x_{2})\log\l1-x_{2}\r+\notag\\
&\frac{x_{2}^2}{1-x_{2}} \log\bl\frac{1}{2} x_{2}^2 (1+x_{2})\br\bigg)\bigg)-\frac{1}{\epsilon^2}\frac{(1+x_{2})^2+2 x_{1} (1+x_{2})^2+x_{1}^2 (1+2 x_{2} (1+x_{2}))}{(1+x_{1}) (1+x_{2})}-\notag\\
&\frac{1}{24\epsilon}\bigg(-\bigg(24 \bigg(-\bigg((1+x_{2}-\mathcal{D}_{0}(1-x_{2})) \mathcal{D}_{1}(1-x_{1})+\notag\\
&2(1+x_{1}-\mathcal{D}_{0}(1-x_{1})) \mathcal{D}_{1}(1-x_{2})+\mathcal{D}_{0}(1-x_{2}) \bigg((1+x_{1}) \log\l1-x_{1}\r+\notag\\
&\frac{x_{1}^2}{1-x_{1}} \log\bl\frac{1}{4} x_{1} (1+x_{1})^2\br\bigg)\bigg)-2(1+x_{2}) \mathcal{D}_{0}(1-x_{1}) \log\l1-x_{2}\r-\notag\\
&\frac{x_{2}^2}{1-x_{2}} \mathcal{D}_{0}(1-x_{1}) \log\bl\frac{1}{2} x_{2}^2 (1+x_{2})\br\bigg)-\delta(1-x_{2}) \bigg(-\pi ^2 (1+x_{1})+\notag\\
&\pi ^2\mathcal{D}_{0}(1-x_{1})+6 \frac{\log^2\l1-x_{1}\r}{1-x_{1}}+6 \bigg(-\mathcal{D}_{2}(1-x_{1})-\frac{x_{1}^2}{1-x_{1}}\bigg(\log^2\l4\r+\notag\\
&2 \log\l4\r \log\bl\frac{1-x_{1}}{x_{1}}\br+\log^2\l1-x_{1}\r-2 \log\l1-x_{1}\r \log\l x_{1}\r-2 \log\bl4 \bigg(\frac{1-x_{1}}{x_{1}}\bigg)\br \log\bl(1+x_{1})^2\br+\notag\\
&\log^2\bl(1+x_{1})^2\br-2 {\rm Li}_{2}\l1-x_{1}\r\bigg)\bigg)\bigg)\bigg)+2 \delta(1-x_{1})\bigg(-\pi ^2(1+x_{2})+\pi ^2\mathcal{D}_{0}(1-x_{2})-\notag\\
&24 \mathcal{D}_{2}(1-x_{2})+24\frac{\log^2\l1-x_{2}\r}{1-x_{2}}-6 \frac{x_{2}}{1-x_{2}}\bigg(x_{2} \bigg(4 \log\l1-x_{2}\r \log\bl\frac{2 (1-x_{2})}{x_{2}^2 (1+x_{2})}\br+\notag\\
&\log^2\bl\frac{1}{2} x_{2} (1+x_{2})\br+\log\l x_{2}\r \log\bl\frac{1}{4} x_{2} (1-x_{2}^2)^2\br\bigg)\bigg)-14\delta(1-x_{2}) \zeta_{3}\bigg)\bigg)+\notag\\
&\frac{1}{\epsilon}\bigg(((1+x_{2})^2+2 x_{1} (1+x_{2})^2+x_{1}^2 (1+2 x_{2} (1+x_{2}))) \frac{1}{(1+x_{1}) (1+x_{2})}\log\l1-x_{1}\r+\notag\\
&2 ((1+x_{2})^2+2 x_{1} (1+x_{2})^2+x_{1}^2 (1+2 x_{2} (1+x_{2}))) \frac{1}{(1+x_{1}) (1+x_{2})}\log\l1-x_{2}\r+\notag\\
&\frac{1}{(1-x_{1})(1-x_{2})} \bigg(x_{1}^2 \log\bl\frac{1}{4} x_{1} (1+x_{1})^2\br+x_{2}^2 \log\bl\frac{1}{2} x_{2}^2 (1+x_{2})\br\bigg)-\notag\\
&2\frac{x_{1}^2 x_{2}^2 (1+x_{1} x_{2})}{(1-x_{1}^2)(1-x_{2}^2)} \log\bl\frac{x_{2}^2 (x_{1}+x_{2})^4}{(1+x_{1}) (1+x_{2})^2 (1+x_{1} x_{2})}\br-\frac{2 x_{1}^2 x_{2}^2 (1+x_{1} x_{2})}{(1-x_{1}^2)(1-x_{2}^2)}\log\bl\frac{x_{1} (1+x_{1} x_{2})}{x_{1}+x_{2}}\br\bigg)+\notag\\
&\frac{1}{24} \bigg(12 (1+x_{2}) \mathcal{D}_{2}(1-x_{1})-12 (4 \mathcal{D}_{1}(1-x_{1}) \mathcal{D}_{1}(1-x_{2})+\mathcal{D}_{0}(1-x_{2}) \mathcal{D}_{2}(1-x_{1}))+\notag\\
&48 (1+x_{1}) \mathcal{D}_{2}(1-x_{2})+\delta(1-x_{2}) \bigg(\pi ^2 \mathcal{D}_{1}(1-x_{1})-2 \mathcal{D}_{3}(1-x_{1})\bigg)+48 \mathcal{D}_{1}(1-x_{2}) \log\l1-x_{1}\r+\notag\\
&48 x_{1} \mathcal{D}_{1}(1-x_{2}) \log\l1-x_{1}\r+\frac{48 x_{1}^2}{1-x_{1}}\mathcal{D}_{1}(1-x_{2}) \log\bl\frac{1}{4} x_{1} (1+x_{1})^2\br+48 \mathcal{D}_{1}(1-x_{1}) \log\l1-x_{2}\r+\notag\\
&48 x_{2} \mathcal{D}_{1}(1-x_{1}) \log\l1-x_{2}\r+\frac{24 x_{2}^2}{1-x_{2}}\mathcal{D}_{1}(1-x_{1}) \log\bl\frac{1}{2} x_{2}^2 (1+x_{2})\br-\frac{2}{1-x_{2}} \mathcal{D}_{0}(1-x_{1}) \bigg(\pi ^2 (1-x_{2}^2)-\notag\\
&24 (1-x_{2}^2) \log^2\l1-x_{2}\r-24 x_{2}^2 \log\l1-x_{2}\r \log\bl\frac{1}{2} x_{2}^2 (1+x_{2})\br+6 x_{2}^2 \bigg(\log^2\bl\frac{1}{2} x_{2} (1+x_{2})\br+\notag\\
&\log\l x_{2}\r \log\bl\frac{1}{4} x_{2} (1-x_{2}^2)^2\br\bigg)\bigg)+\frac{2}{1-x_{1}} \mathcal{D}_{0}(1-x_{2}) \bigg(-\pi ^2(1-x_{1}^2)+6(1-x_{1}^2) \log^2\l1-x_{1}\r+\notag\\
&12 x_{1}^2 \log\l1-x_{1}\r \log\l x_{1}\r+6 x_{1}^2 \log\bl\frac{4 (1-x_{1})^2}{x_{1}^2 (1+x_{1})^2}\br \log\bl\frac{1}{4} (1+x_{1})^2\br+12 x_{1}^2 {\rm Li}_{2}\l1-x_{1}\r\bigg)-\notag\\
&\frac{1}{1-x_{2}}2 \delta(1-x_{1}) \bigg(-16(1-x_{2}^2) \log^3\l1-x_{2}\r-24 x_{2}^2 \log^2\l1-x_{2}\r\log\bl\frac{1}{2} x_{2}^2 (1+x_{2})\br+\notag\\
&2 \log\l1-x_{2}\r \bigg(\pi ^2(1-x_{2}^2)+6 x_{2}^2 \bigg(\log^2\bl\frac{1}{2} x_{2} (1+x_{2})\br+\log\l x_{2}\r \log\bl\frac{1}{4} x_{2}(1-x_{2}^2)^2\br\bigg)\bigg)+\notag\\
&x_{2}^2 \bigg(\pi ^2 \log\bl\frac{1}{2} x_{2} (1+x_{2})\br-2 \log^3\bl\frac{1}{2} x_{2} (1+x_{2})\br-7 \pi ^2 \log\bl\frac{1}{2}(1-x_{2}^2)\br+2\log^3\bl\frac{1}{2}(1-x_{2}^2)\br+\notag\\
&7 \pi ^2 \log\bl\frac{1}{2}x_{2}(1-x_{2}^2)\br-2 \log^3\bl\frac{1}{2}x_{2}(1-x_{2}^2)\br\bigg)-28(1-x_{2}^2) \zeta_{3}\bigg)+2 \mathcal{D}_{0}(1-x_{1}) \bigg(\pi ^2 \mathcal{D}_{0}(1-x_{2})-\notag\\
&24 \mathcal{D}_{2}(1-x_{2})-14 \delta(1-x_{2}) \zeta_{3}\bigg)-\frac{1}{1-x_{1}}\delta(1-x_{2}) \bigg(2 x_{1}^2 \log^3\l4\r+\log\l1-x_{1}\r \bigg(\pi ^2+\notag\\
&2 \log\l1-x_{1}\r (x_{1}^2 \log\l64\r-(1-x_{1}^2) \log\l1-x_{1}\r)-6 x_{1}^2 \log\l16 (1-x_{1})\r \log\l x_{1}\r\bigg)+\notag\\
&6 x_{1}^2 \log\l4\r \log\bl\frac{1-x_{1}}{x_{1}}\br\log\l\frac{4}{(1+x_{1})^4}\r+x_{1}^2 \bigg(\pi ^2-6 \log^2\l4\r-6 \log\l1-x_{1}\r \log\bl\frac{1-x_{1}}{x_{1}^2}\br\bigg) \notag\\
&\times\log\bl(1+x_{1})^2\br-2 x_{1}^2 \log^3\bl(1+x_{1})^2\br-x_{1}^2 \log\bl4\frac{1-x_{1}}{x_{1}}\br(\pi ^2-6 \log^2\l(1+x_{1})^2\r)-\notag\\
&28 \zeta_{3}+4 x_{1}^2 \bigg(3 \log\bl\frac{(1+x_{1})^2}{4 (1-x_{1})}\br {\rm Li}_{2}\l1-x_{1}\r+3 {\rm Li}_{3}\l1-x_{1}\r+7 \zeta_{3}\bigg)\bigg)+\frac{1}{120} \delta(1-x_{1}) (47 \pi ^4 \delta(1-x_{2})+\notag\\
&480(\pi ^2 \mathcal{D}_{1}(1-x_{2})-8 \mathcal{D}_{3}(1-x_{2})-14 \mathcal{D}_{0}(1-x_{2}) \zeta_{3}))\bigg)+\frac{2 x_{1}^2}{(1-x_{1})(1-x_{2})} \log\l x_{1}\r \log\bl\frac{1+x_{1}}{2}\br-\notag\\
&((1+x_{2})^2+2 x_{1} (1+x_{2})^2+x_{1}^2 (1+2 x_{2} (1+x_{2}))) \frac{\log^2\l1-x_{1}\r}{2 (1+x_{1}) (1+x_{2})}-\notag\\
&2((1+x_{2})^2+2 x_{1} (1+x_{2})^2+x_{1}^2 (1+2 x_{2} (1+x_{2})))\frac{\log^2\l1-x_{2}\r}{(1+x_{1}) (1+x_{2})}-\notag\\
&\frac{\log\l1-x_{1}\r}{(1-x_{1}^2)(1-x_{2}^2)} \bigg((1+x_{1}) (1+x_{2}) \bigg(x_{1}^2 \log\bl\frac{1}{4} x_{1} (1+x_{1})^2\br+x_{2}^2 \log\bl\frac{1}{2} x_{2}^2 (1+x_{2})\br\bigg)-\notag\\
&2 x_{1}^2 x_{2}^2 (1+x_{1} x_{2}) \log\bl\frac{x_{2}^2 (x_{1}+x_{2})^4}{(1+x_{1}) (1+x_{2})^2 (1+x_{1} x_{2})}\br\bigg)+\notag\\
&\log\l1-x_{2}\r \bigg(-2 \bigg((1+x_{2})^2+2 x_{1} (1+x_{2})^2+x_{1}^2 (1+2 x_{2} (1+x_{2}))\bigg) \frac{\log\l1-x_{1}\r}{(1+x_{1}) (1+x_{2})}-\notag\\
&\frac{2}{(1-x_{1}^2)(1-x_{2}^2)} \bigg((1+x_{1}) (1+x_{2}) \bigg(x_{1}^2 \log\bl\frac{1}{4} x_{1} (1+x_{1})^2\br+x_{2}^2 \log\bl\frac{1}{2} x_{2}^2 (1+x_{2})\br\bigg)-\notag\\
&2 x_{1}^2 x_{2}^2 (1+x_{1} x_{2}) \log\bl\frac{x_{2}^2 (x_{1}+x_{2})^4}{(1+x_{1}) (1+x_{2})^2 (1+x_{1} x_{2})}\br\bigg)\bigg)+\bigg(\frac{2 x_{1}^2 x_{2}^2 (1+x_{1} x_{2}) \log\l1-x_{1}\r}{(1-x_{1}^2)(1-x_{2}^2)}+\notag\\
&\frac{4 x_{1}^2 x_{2}^2 (1+x_{1} x_{2}) \log\l1-x_{2}\r}{(1-x_{1}^2)(1-x_{2}^2)}-\frac{2 x_{1}^2 x_{2}^2 (1+x_{1} x_{2})}{(1-x_{1}^2)(1-x_{2}^2)}\log\bl\frac{x_{2} (x_{1}+x_{2})^3}{(1+x_{1}) (1+x_{2})^2}\br\bigg) \log\bl\frac{x_{1} (1+x_{1} x_{2})}{x_{1}+x_{2}}\br+\notag\\
&\frac{1}{12(1-x_{1}^2)(1-x_{2}^2)}\bigg(\pi ^2 (1-x_{1})(1-x_{2}) \bigg((1+x_{2})^2+2 x_{1} (1+x_{2})^2+x_{1}^2 (1+2 x_{2} (1+x_{2}))\bigg)+\notag\\
&6 (1+x_{1}) (1+x_{2}) \bigg(4 x_{1}^2+x_{2}^2\bigg) \log^2\l2\r-48 x_{1}^2 (1+x_{1}) (1+x_{2}) \log\l2\r \log\l1+x_{1}\r+\notag\\
&24 x_{1}^2 (1+x_{1}) (1+x_{2}) \log^2\l1+x_{1}\r+6 x_{2}^2 \bigg(-2 (1+x_{1}) (1+x_{2}) \log\l2\r \log\l x_{2} (1+x_{2})\r+\notag\\
&(1+x_{1}) (1+x_{2}) \log^2\l x_{2}(1+x_{2})\r-2 x_{1}^2 (1+x_{1} x_{2}) \bigg(\log^2\bl\frac{x_{2} (x_{1}+x_{2})^3}{(1+x_{1}) (1+x_{2})^2}\br+\notag\\
&\log^2\bl\frac{x_{2} (x_{1}+x_{2})^3 (1-x_{1} x_{2})}{(1+x_{1}) (1+x_{2})^2}\br\bigg)-(1+x_{1}) (1+x_{2}) \log^2\bl\frac{1}{2}(1-x_{2}^2)\br+\notag\\
&2 x_{1}^2 (1+x_{1} x_{2}) \log^2\bl\frac{(x_{1}+x_{2})^2(1-x_{1}^2 x_{2}^2)}{(1+x_{1})(1+x_{2})^2}\br+(1+x_{1}) (1+x_{2}) \log^2\bl\frac{1}{2}(x_{2}-x_{2}^3)\br\bigg)+\notag\\
&12 x_{1}^2 \bigg(-(1+x_{1}) (1+x_{2}){\rm Li}_{2}\l1-x_{1}\r+2 x_{2}^2 (1+x_{1} x_{2}) \bigg(-{\rm Li}_{2}\bl\frac{(1-x_{1}^2) x_{2}}{(x_{1}+x_{2}) (1-x_{1} x_{2})}\br+\notag\\
&{\rm Li}_{2}\bl\frac{x_{2}-x_{1}^2 x_{2}}{x_{1}+x_{2}}\br+{\rm Li}_{2}\bl\frac{(1-x_{1}^2)x_{2}^2}{1-x_{1}^2 x_{2}^2}\br\bigg)\bigg)\bigg)
%#o####
+\mathcal{O}(\epsilon),
\end{align}

\;
\;

\small
\begin{align}
&(Q^2)^{2+\epsilon}{\rm M_{box}}(s_{12},s_{1j},\frac{s_{1j}}{s_{2j}}) =
%#i####
-\frac{x_{2}^2}{\epsilon^3} \delta(1-x_{1})+\frac{x_{2}^2}{\epsilon^2}\bigg(\mathcal{D}_{0}(1-x_{1})-\delta(1-x_{1}) \log\bl\frac{x_{2}^2 (1+x_{2})}{2(1-x_{2})^2}\br\bigg)+\notag\\
&\frac{x_{2}^2 }{\epsilon^2}\bigg(-2+\frac{1}{1+x_{1}}-\frac{2 x_{1}^2 x_{2}}{x_{1}+x_{2}}\bigg)+\frac{1}{12\epsilon}x_{2}^2 \bigg(-12 \bigg(\mathcal{D}_{1}(1-x_{1})+\mathcal{D}_{0}(1-x_{1}) \log\bl\frac{2 (1-x_{2})^2}{x_{2}^2 (1+x_{2})}\br\bigg)+\notag\\
&\delta(1-x_{1}) \bigg(\pi ^2-6 \log^2\bl\frac{x_{2} (1+x_{2})}{2 (1-x_{2})^2}\br+6 \log\bl\frac{1}{x_{2}}\br \log\bl\frac{x_{2} (1+x_{2})^2}{4(1-x_{2})^2}\br\bigg)\bigg)+\notag\\
&\frac{1}{\epsilon}\bigg(x_{2}^2 \big(x_{2}+x_{1}(1+2 x_{1}+2(1+x_{1}+x_{1}^2)x_{2})\big)\frac{\log\l1-x_{1}\r}{(1+x_{1}) (x_{1}+x_{2})}-\frac{x_{2}^2}{1-x_{1}}\bigg(\log\bl\frac{x_{2}^2 (1+x_{2})}{2 (-1+x_{2})^2}\br-\notag\\
&2 \frac{x_{1}^3 (1+x_{1} x_{2})}{(1+x_{1})(x_{1}+x_{2})} \log\bl\frac{x_{1} x_{2}^2 (x_{1}+x_{2})^3}{(1+x_{1}) (1-x_{2}^2)^2}\br\bigg)\bigg)+\notag\\
&\frac{1}{12} x_{2}^2 \bigg(6 \mathcal{D}_{2}(1-x_{1})-12 \mathcal{D}_{1}(1-x_{1}) \log\bl\frac{x_{2}^2 (1+x_{2})}{2(1-x_{2})^2}\br-\notag\\
&\mathcal{D}_{0}(1-x_{1}) \bigg(\pi ^2-6 \log^2\bl\frac{x_{2} (1+x_{2})}{2(1-x_{2})^2}\br+6 \log\bl\frac{1}{x_{2}}\br \log\bl\frac{x_{2} (1+x_{2})^2}{4(1-x_{2})^2}\br\bigg)+\notag\\
&\delta(1-x_{1}) \bigg(-7 \pi ^2 \log\bl\frac{1+x_{2}}{2(1-x_{2})}\br+2 \log^3\bl\frac{1+x_{2}}{2(1-x_{2})}\br+7 \pi ^2 \log\bl\frac{x_{2} (1+x_{2})}{2(1-x_{2})}\br-\notag\\
&2 \log^3\bl\frac{x_{2} (1+x_{2})}{2(1-x_{2})}\br+\pi ^2 \log\bl\frac{x_{2} (1+x_{2})}{2(1-x_{2})^2}\br-2 \log^3\bl\frac{x_{2} (1+x_{2})}{2(1-x_{2})^2}\br+28 \zeta_{3}\bigg)\bigg)-\notag\\
&x_{2}^2 \big(x_{2}+x_{1}(1+2 x_{1}+2(1+x_{1}+x_{1}^2) x_{2})\big) \frac{\log^2\l1-x_{1}\r}{2 (1+x_{1}) (x_{1}+x_{2})}+\frac{2 x_{1}^3 x_{2}^2 (1+x_{1} x_{2})}{(1-x_{1}^2)(x_{1}+x_{2})}\notag\\
&\times\log\bl\frac{x_{1} (1+x_{1} x_{2})}{x_{1}+x_{2}}\br\log\bl\frac{x_{2} (x_{1}+x_{2})^3}{(1+x_{1})(1-x_{2}^2)^2}\br+\frac{x_{2}^2}{1-x_{1}}\log\l1-x_{1}\r\bigg(\log\bl\frac{x_{2}^2 (1+x_{2})}{2 (1-x_{2})^2}\br-\notag\\
&2 \frac{x_{1}^3 (1+x_{1} x_{2})}{(1+x_{1})(x_{1}+x_{2})} \log\bl\frac{x_{1} x_{2}^2 (x_{1}+x_{2})^3}{(1+x_{1})(1-x_{2}^2)^2}\br\bigg)-\frac{1}{12(1-x_{1}^2)(x_{1}+x_{2})}x_{2}^2 \bigg(-\pi ^2 (1-x_{1}) \big(x_{2}+\notag\\
&x_{1}(1+2 x_{1}+2(1+x_{1}+x_{1}^2) x_{2})\big)-6 (1+x_{1}) (x_{1}+x_{2}) \log^2\bl\frac{1+x_{2}}{2-2 x_{2}}\br+\notag\\
&6 (1+x_{1}) (x_{1}+x_{2}) \log^2\bl\frac{x_{2} (1+x_{2})}{2-2 x_{2}}\br+6 \bigg((1+x_{1}) (x_{1}+x_{2}) \log^2\bl\frac{x_{2} (1+x_{2})}{2 (-1+x_{2})^2}\br-\notag\\
&2 x_{1}^3 (1+x_{1} x_{2}) \bigg(\log^2\bl\frac{x_{2} (x_{1}+x_{2})^3}{(1+x_{1})(1-x_{2}^2)^2}\br+\log^2\bl\frac{x_{2} (x_{1}+x_{2})^3 (1-x_{1} x_{2})}{(1+x_{1})(1-x_{2}^2)^2}\br-\notag\\
&\log^2\bl\frac{(x_{1}+x_{2})^2(1-x_{1}^2 x_{2}^2)}{(1+x_{1})(1-x_{2}^2)^2}\br\bigg)\bigg)+24 x_{1}^3 (1+x_{1} x_{2}) \bigg(-{\rm Li}_{2}\bl\frac{(1-x_{1}^2) x_{2}}{(x_{1}+x_{2})(1-x_{1} x_{2})}\br+\notag\\
&{\rm Li}_{2}\bl\frac{x_{2}-x_{1}^2 x_{2}}{x_{1}+x_{2}}\br+{\rm Li}_{2}\bl\frac{(1-x_{1}^2)x_{2}^2}{1-x_{1}^2 x_{2}^2}\br\bigg)\bigg)
%#o####
+\mathcal{O}(\epsilon),
\end{align}

\;
\;

\small
\begin{align}
&(Q^2)^{2+\epsilon}{\rm M_{box}}(s_{1j},s_{2j},\frac{s_{1j}}{s_{12}}) =
%#i####
-\frac{1}{2\epsilon^3}x_{2}^2 \delta(1-x_{1})+\frac{1}{2\epsilon^2}x_{2}^2 \bigg(2 \mathcal{D}_{0}(1-x_{1})+\delta(1-x_{1}) \log\bl\frac{4 (1-x_{2})^2}{x_{2} (1+x_{2})^2}\br\bigg)-\notag\\
&\frac{1}{\epsilon^2} \frac{x_{2}^2}{(1+x_{1}) (x_{1}+x_{2})}\left(x_{2}+x_{1}(1+2 x_{1}+2(1+x_{1}+x_{1}^2)x_{2})\right)-\notag\\
&\frac{x_{2}^2}{\epsilon}\bigg(2 \mathcal{D}_{1}(1-x_{1})+\mathcal{D}_{0}(1-x_{1}) \log\bl\frac{4 (1-x_{2})^2}{x_{2} (1+x_{2})^2}\br\bigg)+\frac{1}{24\epsilon}x_{2}^2 \delta(1-x_{1}) \bigg(5 \pi ^2+6 \log^2\bl\frac{4}{(1+x_{2})^2}\br-\notag\\
&12 \log^2\bl\frac{1}{4} (1+x_{2})^2\br+12 \log\bl\frac{(1-x_{2})^2}{x_{2}}\br \log\bl\frac{(1+x_{2})^2}{4 (1-x_{2})}\br-12 {\rm Li}_{2}\l1-x_{2}\r\bigg)+ \notag\\
&\frac{1}{\epsilon}\bigg(2 x_{2}^2 \big(x_{2}+x_{1}(1+2 x_{1}+2(1+x_{1}+x_{1}^2) x_{2})\big) \frac{\log\l(1-x_{1}) (1-x_{2})\r}{(1+x_{1}) (x_{1}+x_{2})}-\frac{x_{2}^2}{1-x_{1}} \log\bl\frac{4 x_{2}}{(1+x_{2})^2}\br-\notag\\
&\frac{4 (1+x_{1}) x_{2}^2}{1-x_{1}^2}\log\bl\frac{1+x_{2}}{2}\br+\frac{2 x_{1}^3 x_{2}^2 (1+x_{1} x_{2})}{(1-x_{1}^2) (x_{1}+x_{2})} \log\bl\frac{x_{1} x_{2} (x_{1}+x_{2})^4}{(1+x_{1})^2 (1+x_{2})^2}\br\bigg)+\notag\\
&\frac{1}{24}x_{2}^2 \bigg(48 \mathcal{D}_{2}(1-x_{1})+48 \mathcal{D}_{1}(1-x_{1}) \log\bl\frac{4 (1-x_{2})^2}{x_{2} (1+x_{2})^2}\br+2 \mathcal{D}_{0}(1-x_{1}) \bigg(-5 \pi ^2-6 \log^2\bl\frac{4}{(1+x_{2})^2}\br+\notag\\
&24 \log\l1-x_{2}\r \log\bl\frac{4 (1-x_{2})}{(1+x_{2})^2}\br+12 \log^2\bl\frac{1}{4} (1+x_{2})^2\br+12 \log\l x_{2}\r \log\bl\frac{(1+x_{2})^2}{4 (1-x_{2})}\br+\notag\\
&12 {\rm Li}_{2}\l1-x_{2}\r\bigg)+\delta(1-x_{1}) \bigg(16 \log^3\l1-x_{2}\r+5 \pi ^2 \log\bl\frac{4}{(1+x_{2})^2}\br-2 \log^3\bl\frac{4}{(1+x_{2})^2}\br+\notag\\
&10 \pi ^2 \log\bl\frac{1}{4} (1+x_{2})^2\br-24 \log^2\l1-x_{2}\r \log\bl\frac{1}{4} (1+x_{2})^2\br-4 \log^3\bl\frac{1}{4} (1+x_{2})^2\br-\notag\\
&2 \log\l1-x_{2}\r \bigg(5 \pi ^2+6 \log^2\bl\frac{4}{(1+x_{2})^2}\br-12 \log^2\bl\frac{1}{4} (1+x_{2})^2\br\bigg)+\log\l x_{2}\r \bigg(\pi ^2-\notag\\
&6 \log^2\bl\frac{(1+x_{2})^2}{4(1-x_{2})}\br\bigg)-12 \log\bl\frac{(1+x_{2})^2}{4(1-x_{2})}\br {\rm Li}_{2}\l1-x_{2}\r+12{\rm Li}_{3}\l1-x_{2}\r+28 \zeta_{3}\bigg)\bigg)-\notag\\
&\frac{(x_{1}^3+x_{1}^3 x_{2}-(1-x_{1}) x_{2}^3) \log^2\l4\r}{(1-x_{1}) (x_{1}+x_{2})}+\frac{1}{(1-x_{1}) (x_{1}+x_{2})}2 \bigg(2 (x_{1}^3+x_{1}^3 x_{2}+(-1+x_{1}) x_{2}^3) \log^2\l2\r+\notag\\
&x_{1} x_{2}^2 (1+x_{2}) \log\bl\frac{1}{4} (1+x_{1})^2\br \log\l(1-x_{1}) (1-x_{2})\r+\log\l1-x_{2}\r \bigg(-x_{1} (1-x_{2}) x_{2}^2 \log\l4\r-\notag\\
&4 x_{1}^3 (1+x_{2}) \log\l1+x_{1}\r+x_{1}^3 (1+x_{2}) \log\bl4 (1+x_{1})^2\br+x_{1}^3 \log\bl\frac{1}{4} (1+x_{1})^2\br \log\l(1-x_{1})(1-x_{2})\r+\notag\\
&x_{1}^3 x_{2} \log\bl\frac{1}{4} (1+x_{1})^2\br \log\l(1-x_{1})(1-x_{2})\r-x_{1} x_{2}^2 \log\bl\frac{1}{16} (1+x_{2})^2\br-\notag\\
&x_{1} x_{2}^3 \log\bl(1+x_{2})^2\br\bigg)+x_{1} \log\l1-x_{1}\r \bigg(-x_{1}^2 (1+x_{1}) (1+x_{2}) \log\bl\frac{1}{4} (1+x_{1})^2\br+\notag\\
&x_{2}^2 \bigg(x_{1} \log\l16\r-4 (1+x_{1}) (1+x_{2}) \log\l1+x_{2}\r+\log\bl4 (1+x_{2})^2\br+x_{2} \log\bl64 (1+x_{2})^2\br-\notag\\
&(1-x_{1}) x_{2} \log\bl4 x_{2} (1+x_{2})^2\br\bigg)\bigg)\bigg)-\frac{2 x_{1}^3 x_{2}^2 (1+x_{1} x_{2})}{(1-x_{1}^2) (x_{1}+x_{2})} \bigg(\log\bl\frac{1}{(1-x_{2})^2}\br \log\bl\frac{x_{1} (1+x_{1} x_{2})}{x_{1}+x_{2}}\br+\notag\\
&\log\bl\frac{1}{(1-x_{1})^2}\br \log\bl\frac{x_{2} (1+x_{1} x_{2})}{x_{1}+x_{2}}\br-2 \bigg(\log\l1-x_{2}\r \log\bl\frac{x_{1}+x_{2}}{x_{1}+x_{1}^2 x_{2}}\br+\log\l1-x_{1}\r \notag\\
&\times\log\bl\frac{x_{1}+x_{2}}{x_{2}+x_{1} x_{2}^2}\br\bigg)\bigg)-2 x_{2}^2 \big(x_{2}+x_{1}(1+2 x_{1}+2(1+x_{1}+x_{1}^2) x_{2})\big) \frac{\log^2\l1-x_{1}\r+\log^2\l1-x_{2}\r}{(1+x_{1}) (x_{1}+x_{2})}+\notag\\
&\frac{\log\l(1-x_{1}) (1-x_{2})\r}{(1-x_{1}) (x_{1}+x_{2})}\log\bl\frac{4}{(1+x_{1})^2}\br\big(2 x_{1}^3 (1+x_{2})\log\l1-x_{2}\r +2 x_{1} x_{2}^2 (1+x_{2})\big)+\notag\\
&\frac{x_{2}^2 }{2-2 x_{1}}\log^2\bl\frac{4}{(1+x_{2})^2}\br-\frac{x_{2}^2 \log\l x_{2}\r }{1-x_{1}}\log\bl\frac{(1+x_{2})^2}{4 (1-x_{2})}\br+\frac{2 x_{1}^3 x_{2}^2 (1+x_{1} x_{2})}{(1-x_{1}^2)(x_{1}+x_{2})} \log^2\bl\frac{(x_{1}+x_{2})^4}{(1+x_{1})^2 (1+x_{2})^2}\br-\notag\\
&8 x_{1}^3 x_{2}^2 (1+x_{1} x_{2}) \log\l(1-x_{1}) (1-x_{2})\r\frac{1}{(1-x_{1}^2)(x_{1}+x_{2})}\log\bl\frac{(x_{1}+x_{2})^4}{(1+x_{1})^2 (1+x_{2})^2}\br+\notag\\
&\log\l1-x_{2}\r \bigg(-4 x_{2}^2 \big(x_{2}+x_{1}(1+2 x_{1}+2(1+x_{1}+x_{1}^2)x_{2})\big) \frac{\log\l1-x_{1}\r}{(1+x_{1}) (x_{1}+x_{2})}+\frac{2 x_{1}^3 (1+x_{2})\log\l x_{1}\r}{(1-x_{1})(x_{1}+x_{2})}-\notag\\
&\frac{2 }{(1-x_{1}) (x_{1}+x_{2})}\bigg(-4 x_{1}^3 (1+x_{2}) \log\l1+x_{1}\r+x_{1}^3 (1+x_{2}) \log\bl4 x_{1} (1+x_{1})^2\br+\notag\\
&x_{2}^2 \bigg(x_{1} \log\l16\r-4 (x_{1}+x_{2}) \log\l1+x_{2}\r+x_{2} \big(x_{1} \log\bl\frac{4}{(1+x_{2})^2}\br+\log\bl4 (1+x_{2})^2\br\big)\bigg)\bigg)-\notag\\
&4 x_{1}^3 x_{2}^2 \frac{(1+x_{1} x_{2})}{(1-x_{1}^2)(x_{1}+x_{2})} \log\bl\frac{(1+x_{1})^2 (1+x_{2})^2}{(x_{1}+x_{2})^2 (1+x_{1} x_{2})^2}\br\bigg)+\log\l1-x_{1}\r \bigg(\frac{2 x_{1} x_{2}^2 (1+x_{2}) \log\l x_{2}\r}{(1-x_{1}) (x_{1}+x_{2})}-\notag\\
&2 \frac{(1+x_{1})}{(1-x_{1}) (x_{1}+x_{2})} \bigg(x_{1}^3 (1+x_{2}) \log\bl\frac{4}{(1+x_{1})^2}\br+x_{2}^2 \bigg(x_{1} \log\l16\r-4 x_{1} (1+x_{2}) \log\l1+x_{2}\r-\notag\\
&x_{2} \log\bl\frac{1}{4} x_{2} (1+x_{2})^2\br+x_{1} x_{2} \log\bl4 x_{2} (1+x_{2})^2\br\bigg)\bigg)-4 x_{1}^3 x_{2}^2 \frac{(1+x_{1} x_{2}) }{(1-x_{1}^2)(x_{1}+x_{2})}\notag\\
&\times\log\bl\frac{(1+x_{1})^2 (1+x_{2})^2}{(x_{1}+x_{2})^2 (1+x_{1} x_{2})^2}\br\bigg)+\frac{2 x_{1}^3 x_{2}^2 (1+x_{1} x_{2})}{(1-x_{1}^2)(x_{1}+x_{2})} \log\bl\frac{x_{2}(x_{1}+x_{2})^3}{(1-x_{1}^2)(1-x_{2}^2)^2}\br \log\bl\frac{x_{1} (1+x_{1} x_{2})}{x_{1}+x_{2}}\br+\notag\\
&\frac{2 x_{1}^3 x_{2}^2 (1+x_{1} x_{2})}{(1-x_{1}^2)(x_{1}+x_{2})} \log\bl\frac{x_{2} (1+x_{1} x_{2})}{x_{1}+x_{2}}\br \log\bl\frac{x_{1}(x_{1}+x_{2})^3}{(1-x_{1}^2)^2(1-x_{2}^2)}\br-\frac{1}{12(1-x_{1}^2)(x_{1}+x_{2})}\notag\\
&\times x_{2}^2 \bigg(\pi ^2 \bigg(-5 x_{2}+x_{1} \big(-5 (1+x_{2})+x_{1} (-5+6 x_{1} (1+x_{1} x_{2}))\big)\bigg)+12 (1+x_{1}) (x_{1}+x_{2}) \log^2\l4\r+\notag\\
&48 (1+x_{1}) (x_{1}+x_{2}) \log\bl\frac{1+x_{2}}{4}\br \log\l1+x_{2}\r+12 x_{1}^3 (1+x_{1} x_{2}) \log^2\bl\frac{(1+x_{1})^2 (1+x_{2})^2}{(x_{1}+x_{2})^2 (1+x_{1} x_{2})^2}\br+\notag\\
&12 (1+x_{1}) (x_{1}+x_{2}) {\rm Li}_{2}\l1-x_{2}\r+24 x_{1}^3 (1+x_{1} x_{2}) \bigg({\rm Li}_{2}\bl\frac{(x_{1}+x_{2})^2}{(1+x_{1} x_{2})^2}\br-{\rm Li}_{2}\bl\frac{x_{2}-x_{1}^2 x_{2}}{x_{1}+x_{2}}\br-\notag\\
&{\rm Li}_{2}\bl\frac{x_{1}-x_{1} x_{2}^2}{x_{1}+x_{2}}\br\bigg)\bigg)
%#o####
+\mathcal{O}(\epsilon).
\end{align}

\noindent The remaining box integrals can be 
obtained from the previous ones by exploiting the symmetry with respect to the transformation $x_{1}\leftrightarrow x_{2}$. More precisely, ${\rm M_{box}}(s_{12},s_{2j},1)$,  
${\rm M_{box}}(s_{1j},s_{2j},s_{2j}/s_{12})$, 
${\rm M_{box}}(s_{12},s_{2j},s_{12}/s_{1j})$ and 
${\rm M_{box}}(s_{12},s_{2j},s_{2j}/s_{1j})$ can be derived from  ${\rm M_{box}}(s_{12},s_{1j},1)$, \\
${\rm M_{box}}(s_{1j},s_{2j},s_{1j}/s_{12})$,  ${\rm M_{box}}(s_{12},s_{1j},s_{12}/s_{2j})$ and 
${\rm M_{box}}(s_{12},s_{1j},s_{1j}/s_{2j})$ respectively by exchanging $x_{1}\leftrightarrow x_{2}$.

\section{Example of an integrated antenna function: $\hat{{\cal D}}_{qg,g}^{1}$ antenna}
\label{app:B}
The integrated one-loop antenna functions ${\cal X}_{ik,j}$ result in 
lengthy expressions. As an example of these, the function $\hat{{\cal D}}_{qg,g}^{1}(x_1,x_2)$
reads:

\small
\begin{align}
&(Q^2)^{2\epsilon}\hat{D}_{qg,g}^{1}(x_1,x_2) = 
%#i####
\frac{1}{3\epsilon^3} \delta(1-x_{1}) \delta(1-x_{2})+\frac{1}{6\epsilon^2} \delta(1-x_{2}) (1+x_{1}-2 \mathcal{D}_{0}(1-x_{1}))+\notag\\
&\frac{1}{3 x_{2}\epsilon^2} \delta(1-x_{1}) (-1+x_{2} (2+(-1+x_{2}) x_{2})-x_{2} \mathcal{D}_{0}(1-x_{2}))+\notag\\
&\frac{1}{36\epsilon} \bigg(6 \bigg(\frac{1}{(1+x_{1}) x_{2} (1+x_{2}) (x_{1}+x_{2})^3}\bigg(2 x_{1}^2 (2+x_{1})+x_{1}\left(4+x_{1}(2+x_{1})(1+x_{1}^2)\right) x_{2}+\notag\\
&\bigg(2+x_{1}^2 \big(3+x_{1} \left(10+x_{1} (7+2 x_{1})\right)\big)\bigg) x_{2}^2+x_{1} \left(9+x_{1} (2+x_{1}) \left(9+2 x_{1} \left(3+x_{1}\right)\right)\right) x_{2}^3+\notag\\
&\bigg(3+x_{1} \big(12+x_{1} \left(19+2 x_{1} \left(7+x_{1} \left(5+x_{1}\right)\right)\right)\big)\bigg) x_{2}^4+\bigg(1+x_{1} \big(1+x_{1} \left(4+x_{1} (2+x_{1})\right)\big)\bigg)\notag\\
&\times2(1+x_{1})x_{2}^5+2 x_{1} (3+2 x_{1} (3+x_{1} (2+x_{1}))) x_{2}^6+2(1+2 x_{1}(1+x_{1}+x_{1}^2)) x_{2}^7\bigg)-\big(2 (-\frac{1}{x_{2}}+\notag\\
&(2-(1-x_{2}) x_{2})) \mathcal{D}_{0}(1-x_{1})+(1+x_{1}-2 \mathcal{D}_{0}(1-x_{1})) \mathcal{D}_{0}(1-x_{2})\big)+\delta(1-x_{2}) \bigg((1-x_{1})+\notag\\
&2\mathcal{D}_{1}(1-x_{1})-(1+x_{1})\log\l1-x_{1}\r+\frac{(1+x_{1}^2)}{1-x_{1}} \log\bl\frac{2}{1+x_{1}}\br\bigg)\bigg)-\notag\\
&\delta(1-x_{1}) \bigg(-\bigg(-\pi ^2\delta(1-x_{2})-3(x_{2}-4 \mathcal{D}_{1}(1-x_{2}))-12 (-\frac{1}{x_{2}}+\notag\\
&(2-(1-x_{2}) x_{2})) \log\l1-x_{2}\r\bigg)-12 \frac{(1-(1-x_{2}) x_{2})^2}{(1-x_{2})x_{2}} \log\bl\frac{2}{1+x_{2}}\br\bigg)\bigg)+\notag\\
&\frac{1}{72} \bigg(2 \pi ^2 \delta(1-x_{2}) \mathcal{D}_{0}(1-x_{1})+\frac{24 (-1+x_{2} (2-(1-x_{2}) x_{2})) \mathcal{D}_{1}(1-x_{1})}{x_{2}}+12 (1+x_{1}) \mathcal{D}_{1}(1-x_{2})-\notag\\
&24 (\mathcal{D}_{0}(1-x_{2}) \mathcal{D}_{1}(1-x_{1})+\mathcal{D}_{0}(1-x_{1}) \mathcal{D}_{1}(1-x_{2}))-12 \delta(1-x_{2}) \mathcal{D}_{2}(1-x_{1})+\notag\\
&2 \delta(1-x_{1}) \bigg(\pi ^2 \mathcal{D}_{0}(1-x_{2})-6 \mathcal{D}_{2}(1-x_{2})\bigg)-\frac{12}{1-x_{1}} \mathcal{D}_{0}(1-x_{2}) \bigg((1-x_{1})^2+x_{1}^2 \log\l2(1-x_{1})\r+\notag\\
&\log\bl\frac{2}{1-x_{1}}\br+(1+x_{1}^2) \log\bl\frac{1}{1+x_{1}}\br\bigg)-\frac{1}{1-x_{1}}\delta(1-x_{2}) \bigg(\pi ^2 (1-x_{1}^2)+6(1+x_{1}^2) \log^2\l2\r+\notag\\
&6(\log\l4\r+x_{1}^2 \log\l4(1-x_{1})\r) \log\l1-x_{1}\r-6 \log^2\l1-x_{1}\r+12 \log\l2(1-x_{1})\r \bigg((1-x_{1})^2+\notag\\
&(1+x_{1}^2)\log\bl\frac{1}{1+x_{1}}\br\bigg)+6 \log\bl\frac{1}{1+x_{1}}\br \bigg(2 (1-x_{1})^2+(1+x_{1}^2) \log\bl\frac{1}{1+x_{1}}\br\bigg)\bigg)+\notag\\
&\frac{12}{x_{2}} \mathcal{D}_{0}(1-x_{1})\bigg(x_{2}^2+2 (-1+x_{2} (2-(1-x_{2}) x_{2})) \log\l1-x_{2}\r\bigg)-\notag\\
&\frac{24\mathcal{D}_{0}(1-x_{1})}{(1-x_{2}) x_{2}}(1-(1-x_{2}) x_{2})^2 \log\bl\frac{2}{1+x_{2}}\br-\frac{1}{(1-x_{2}) x_{2}}2 \delta(1-x_{1}) \bigg(-6(1-x_{2}) (-1+x_{2} (2-\notag\\
&(1-x_{2}) x_{2})) \log^2\l1-x_{2}\r-(-1+x_{2}) \bigg(6+2 x_{2} (-3+7 x_{2})+\pi ^2 (-1+x_{2} (2-(1-x_{2}) x_{2}))-\notag\\
&3 x_{2}^2 \log\bl\frac{4}{(1+x_{2})^2}\br\bigg)+6 (1-(1-x_{2}) x_{2})^2 \log^2\bl\frac{2}{1+x_{2}}\br+3 \log\l1-x_{2}\r \bigg(-(1-x_{2}) x_{2}^2+\notag\\
&4 (1-(1-x_{2}) x_{2})^2 \log\bl\frac{2}{1+x_{2}}\br\bigg)\bigg)+\frac{1}{(1-x_{1}^2) x_{2} (x_{1}+x_{2})^3(1-x_{2}^2)}12 \bigg(x_{2} \bigg(-2 x_{2}+x_{2}^3+\notag\\
&x_{2}^6+x_{1}^6 \bigg(1+x_{2}-2 x_{2}^5\bigg)-x_{1}^2 (1-x_{2}) x_{2} (-1+x_{2} (1+x_{2}) (2+5 x_{2}))-x_{1} (1-x_{2}) x_{2}^2 (5+\notag\\
&x_{2} (6+x_{2} (4+x_{2})))+x_{1}^4 \big(-1+x_{2}-6 x_{2}^5+x_{2}^3 (6+\log\l8\r)\big)-x_{1}^3 \big(-1+x_{2}(3+x_{2}+x_{2}^2-\notag\\
&6 x_{2}^3+2 x_{2}^5-x_{2} \log\l8\r)\big)+x_{1}^5 \big(-1+x_{2}(2-10 x_{2}^3+x_{2} (9+\log\l8\r))\big)\bigg)+\bigg(x_{1}^6 x_{2} (1+x_{2}-2 x_{2}^5)+\notag\\
&x_{1}^5 x_{2} \big(1+x_{2}(4+3 x_{2}-4 x_{2}^3-4 x_{2}^5)\big)-(1-x_{2}) x_{2}^2 \big(2+x_{2}^2(3+2(x_{2}+x_{2}^3))\big)-\notag\\
&x_{1}^3 (1-x_{2})^2 \bigg(-2+x_{2} \big(-3+x_{2}(4+x_{2}(7+2 x_{2}+4 x_{2}^2))\big)\bigg)+x_{1}^2 \bigg(-4+x_{2} \bigg(6+x_{2} \big(-5+\notag\\
&x_{2} (-6+x_{2} (2+x_{2}+6 x_{2}^3))\big)\bigg)\bigg)+x_{1} x_{2} \bigg(-4+x_{2} \big(6+x_{2} (-11+x_{2}^2 (7+2 x_{2} (-2+x_{2} (2+x_{2}))))\big)\bigg)+\notag\\
&x_{1}^4 \bigg(2+x_{2} \bigg(-3+x_{2} \big(4+x_{2} \big(8-x_{2}(7+2 x_{2}(-1+x_{2}+2 x_{2}^3))\big)\big)\bigg)\bigg)\bigg) \log\l1-x_{1}\r+\notag\\
&x_{1}^2 x_{2} \bigg(x_{2}^4+x_{1}^4 (1+x_{2})+x_{1}^3 (1+x_{2}) (1+3 x_{2})+x_{1} x_{2}^2 (3+4 x_{2})+3 x_{1}^2 \bigg(x_{2}+x_{2}^3\bigg)\bigg) \log\bl\frac{1}{1+x_{1}}\br+\notag\\
&x_{2} \bigg(x_{1}^3 (1+x_{1})+x_{1}^2 (3+4 x_{1}) x_{2}+3 x_{1} (1+2 x_{1}) \left(1+x_{1}^2\right) x_{2}^2+(1+2 x_{1})^2 x_{2}^3+\left(1+x_{1}+x_{1}^3\right) x_{2}^4\bigg)\notag\\
&\times\log\bl\frac{2}{1+x_{1}}\br+x_{1}^4 x_{2}^2 \log\bl\frac{16}{1+x_{1}}\br+x_{1}^2 x_{2} \left(x_{1}^3 (1+x_{1})+x_{1}^3 (4+x_{1}) x_{2}+x_{2}^4\right)\log\l2(1-x_{2})\r-\notag\\
&\bigg(-2 x_{1}^6 x_{2}^6+x_{1}^5 x_{2}^3(3-4(x_{2}^2+x_{2}^4))-(1-x_{2}) x_{2}^2 \big(2+x_{2}^2(3+2(x_{2}+x_{2}^3))\big)-\notag\\
&x_{1}^3 (1-x_{2})^2 \bigg(-2+x_{2} \big(-3+x_{2}(4+x_{2}(7+2 x_{2}+4 x_{2}^2))\big)\bigg)+x_{1}^2 \bigg(-4+x_{2} \big(6+x_{2}(-5+2 x_{2}(-3+\notag\\
&x_{2}+3 x_{2}^4))\big)\bigg)+x_{1} x_{2} \bigg(-4+x_{2} \big(6+x_{2} \big(-11+x_{2}^2 (7+2 x_{2} (-2+x_{2} (2+x_{2})))\big)\big)\bigg)+\notag\\
&x_{1}^4 \bigg(2+x_{2} \bigg(-3+x_{2} \big(4+x_{2} \big(8-x_{2} (7+2 x_{2} (-1+x_{2}+2 x_{2}^3))\big)\big)\bigg)\bigg)\bigg) \log\l1-x_{2}\r+\notag\\
&2 \bigg(x_{1}^4 (1+x_{2}) \left(1+(-1+x_{2}) x_{2}\right)^2+x_{2}^3 (1+x_{2}) (1+(-1+x_{2}) x_{2})^2+x_{1} x_{2}^2 (1+x_{2}) (3+x_{2}) \big(1-\notag\\
&(1-x_{2}) x_{2}\big)^2+3 x_{1}^2 x_{2} \left(1+x_{2}^3\right)^2+x_{1}^3 \left(1+x_{2} \left(2+x_{2} \left(-2+4 x_{2}-2 x_{2}^3+3 x_{2}^4\right)\right)\right)\bigg)\log\bl\frac{2}{1+x_{2}}\br+\notag\\
&4 x_{1}^3 x_{2}^4 \log\bl\frac{4}{1+x_{2}}\br-2 (1+x_{1} x_{2}) \bigg(x_{2}^2+x_{1}^5 x_{2}^5+\notag\\
&x_{1} x_{2} (2+x_{2}^2)+x_{1}^3 x_{2}^3 \big(3-x_{2}^2+2 x_{2}^4\big)+x_{1}^2 \big(2-x_{2}^2+3 x_{2}^4\big)+x_{1}^4 \big(x_{2}^4+2 x_{2}^6\big)\bigg) \log\bl\frac{(1+x_{1}) (1+x_{2})}{(x_{1}+x_{2})^2}\br\bigg)-\notag\\
&\frac{1}{(1-x_{1}^2)x_{2}(x_{1}+x_{2})^3(1-x_{2}^2)}24 \bigg(2 x_{1}^6 x_{2}^6+x_{1}^5 x_{2}^3 (-3+4 (x_{2}^2+x_{2}^4))+(1-x_{2}) x_{2}^2 \big(2+x_{2}^2 (3+2 (x_{2}+x_{2}^3))\big)+\notag\\
&x_{1}^3 (1-x_{2})^2 \bigg(-2+x_{2} \big(-3+x_{2}(4+x_{2}(7+2 x_{2}+4 x_{2}^2))\big)\bigg)+x_{1}^2 \bigg(4+x_{2} \big(-6+x_{2} (5-2 x_{2} (-3+\notag\\
&x_{2}+3 x_{2}^4))\big)\bigg)+x_{1} x_{2} \big(4-x_{2} \big(6+x_{2} (-11+x_{2}^2 (7+2 x_{2} (-2+x_{2} (2+x_{2}))))\big)\big)+x_{1}^4 \bigg(-2+x_{2} \bigg(3+\notag\\
&x_{2} \big(-4+x_{2} \big(-8+x_{2} (7+2 x_{2} (-1+x_{2}+2 x_{2}^3))\big)\big)\bigg)\bigg)\bigg) \log\l1-x_{2}\r-8 \delta(1-x_{1}) \delta(1-x_{2}) \zeta_{3}\bigg)
%#o####
+\mathcal{O}(\epsilon).
\end{align}

\end{document}